\documentclass[aps,prb,twocolumn,superscriptaddress,amsmath,showpacs]{revtex4}
\usepackage{color}
\usepackage{graphicx}
\usepackage{bm}
\newcommand{\figwidth}{3.375in}

\begin{document}
\title[Short Title]{
Nature of the phase transition of the three-dimensional isotropic Heisenberg 
spin glass
} 
\author{Koji Hukushima}
\email{hukusima@phys.c.u-tokyo.ac.jp}
\affiliation{
Department of Basic Science, University of Tokyo, 
3-8-1 Komaba, Meguro-ku, Tokyo 153-8902, Japan
} 
\author{Hikaru Kawamura}
\email{kawamura@ess.sci.osaka-u.ac.jp}
\affiliation{
Department  of Earth and Space Science, Faculty of Science, Osaka
University, Toyonaka, Osaka 560-0043, Japan} 

\date{\today}
\begin{abstract}
Equilibrium properties of the three-dimensional
 isotropic Heisenberg spin glass are studied by extensive Monte
 Carlo simulations, with particular attention to the nature of its phase transition. A finite-size-scaling analysis is performed both for the
 spin-glass (SG) and the chiral-glass (CG) orders. 
Our results suggest that the model exhibits the CG long-range 
order at finite
 temperatures 
without accompanying the conventional SG long-range order, in contrast to some of the recent works
 claiming the simultaneous SG and CG transition. 
Typical length and time scales which represent a crossover from
the spin-chirality coupling regime at short scales to the spin-chirality
decoupling regime at long scales are introduced and examined
in order to observe the true asymptotic transition behavior.  On the
basis of these crossover scales, 
discussion is given concerning
the cause of the discrepancy between our present result and those of other
recent numerical works. 
\end{abstract}
\pacs{75.10.Nr, 05.10.Ln,05.70.Fh,64.60.Fr }
\maketitle

\section{Introduction}
\label{sec:introduction}

Spin glasses (SGs) have attracted the attention of researchers both in
experiments and theory as a prototype of complex systems with quenched
randomness.\cite{SGreviews} SGs are random magnets in which magnetic ions
interact with each other either ferromagnetically or antiferromagnetically,
depending on their positions. 
Most of theoretical works have been so far devoted to the minimal SG
model, {\it i.e.\/}, the Ising Edwards and  Anderson (EA) model. 
After a discussion early in the 1980s, it is now widely
believed that a three-dimensional (3D) Ising SG model exhibits  a SG
phase transition at a finite temperature. 
Large-scale Monte Carlo (MC) simulations presented an evidence for
the finite-temperature SG ordering.\cite{BhattYoung, Ogielski}
Subsequently, the critical exponents evaluated by MC
simulations were  consistently compared with 
those evaluated experimentally for the Ising-like SG compound FeMnTiO$_3$. 

Compared to the Ising case,  the nature of the phase transition of 
continuous spin systems such as XY and
Heisenberg SGs are still poorly understood.
Since many SG magnets including canonical SG possess only weak magnetic
anisotropy, an isotropic Heisenberg SG model,  rather than
the strongly anisotropic Ising model, is expected to be a realistic 
model of SG magnets. 
Experimentally, an equilibrium SG phase transition has been established in
real SG materials via measurements of 
the divergent non-linear susceptibility, {\it etc\/}. 
In sharp contrast to experiments, earlier theoretical studies on
the Heisenberg SG model indicated that the standard SG long-range
order occurred only at zero temperature in
three dimensions.\cite{Olive86,Banavar,McMillan,Matsubara0,Yoshino1}

In order to solve this apparent puzzle, a  chirality mechanism of
experimentally observed SG transitions was proposed by
Kawamura\cite{Kawamura92,Kawamura2}. 
This scenario is based on the assumption 
that an isotropic 3D Heisenberg SG exhibits a finite-temperature 
{\it chiral-glass\/} (CG) transition without the conventional SG
order. In terms of symmetry, among the global symmetries of the isotropic
Hamiltonian, only the $Z_2$ spin-reflection (or spin-inversion)
symmetry associated with 
the chirality is
spontaneously broken with keeping the $SO(3)$ spin-rotation symmetry. 
Indeed, some numerical studies\cite{Kawamura3,HukuKawa00a}
claimed that the standard SG order associated with the freezing of the
Heisenberg spin occurred at a temperature lower than the CG transition
temperature, {\it i.e.\/}, $T_{\rm SG}<T_{\rm CG}$, possibly with $T_{{\rm SG}}=0$. 
It means that
the spin and the chirality are \textit{decoupled on long length and
time scales\/}, although the chirality is locally defined as a composite 
operator of the spin variables.

In this chirality scenario of experimental SG transitions, 
essential features of many of the real SG transition and of the 
SG ordered state are
determined by the properties of the CG transition
and of the CG state \textit{of the fully isotropic system}. 
The  role of the magnetic anisotropy  is secondary which
\textit{re-couples} 
the spin to the chirality and \textit{reveals} the CG transition
in the chiral sector
as an anomaly in experimentally accessible spin-related quantities. 
The scenario  successfully explained the
phase diagram under magnetic fields observed by the recent 
numerical simulation \cite{ImagawaKawamura02,ImagawaKawamura04} and 
experiments.\cite{Campbell}

More recently, however, some researchers argued a possibility that in
the 3D Heisenberg SG model the spin ordered 
at a finite temperature simultaneously with the chirality, {\it i.e.\/}, 
$T_{\rm SG}=T_{\rm
CG}>0$.\cite{Matsubara1,Endoh1,Matsubara2,Nakamura,LeeYoung} 
Thus, the nature of the ordering of the 3D Heisenberg SG, as well as the 
validity of the chirality scenario, is now under debate. 
Under such circumstances, it is highly interesting to perform further extensive
numerical studies of the  3D Heisenberg SG in order to clarify the true nature
of its ordering. 
In the present study, we investigate both the SG and
CG orderings of the model by means of a large-scale
equilibrium MC simulation. 

Interestingly, recent experiments reported on a qualitative difference in 
aging phenomena between a canonical Heisenberg-like SG and an Ising-like
SG.\cite{Exp}  We also expect that the full understanding of the equilibrium 
properties of the 3D Heisenberg SG will also give a valuable insight into these
off-equilibrium properties of SGs.

The article is organized as follows. In Sec.~\ref{sec:background}, 
we give a background 
of the present numerical study.
 In Sec.~\ref{sec:basics},
we  explain first the basics of the chirality mechanism. 
In Sec.~\ref{sec:decoupling}, we introduce the 
crossover length and time scales beyond which the spin and the chirality
are decoupled  with each other. These length and time scales are 
crucially important in the chirality
mechanism, and are also essential in properly interpreting  the numerical data 
of MC simulations. 
In Sec.~\ref{sec:model}, we explain the model and the MC method employed. 
In Sec.~\ref{sec:pq}, we introduce various 
physical quantities measured in our MC
simulations, while the results of our simulations 
are presented in Sec.~\ref{sec:NR}. 
In view of our MC results, we examine 
and discuss in Sec.~\ref{sec:discussions} the recent numerical results on the 3D Heisenberg SG by other 
authors.
Finally, we present a brief summary of the results in section VII.

\section{background}
\label{sec:background}

In this section, we wish to give a background of the present numerical study of
the 3D isotropic Heisenberg SG. First, we explain the basics of the 
chirality mechanism of experimental SG transition as proposed in 
Refs.~\onlinecite{Kawamura92} and \onlinecite{Kawamura2}.
Then, we explain the notion of the 
spin-chirality decoupling, together with the crossover length and
time scales which play a crucially important role 
in the chirality mechanism and are also essential in
properly interpreting the numerical data.

\subsection{Chirality mechanism}
\label{sec:basics}

Chirality is an Ising-like multi-spin variable representing the sense or
the handedness of noncollinear spin structures induced by spin
frustration. 
In frustrated magnets with continuous spins, the chirality often plays an
essential role in their magnetic ordering. 
The local chirality $\chi_{i \mu}$ at the $i$th site in the 
$\mu$-direction may be defined by 

\begin{equation}
\chi_{i \mu} = \vec{S}_{i+\hat{e}_\mu} \cdot (\vec{S}_i \times \vec{S}_{i-\hat{e
}_\mu}), 
\label{scalar-chirality}
\end{equation}
$\hat{e}_\mu (\mu =x,y,z)$ being a unit lattice vector along the
$\mu$ axis. 
This quantity 
is often called a scalar chirality: It takes a non-zero value only
when the three neighboring spins take the non-coplanar configuration in
spin space, while it vanishes for the collinear or the coplanar spin
configuration. The chirality defined above is a pseudoscalar variable
since it is invariant under the global $SO(3)$ spin rotations but
changes its sign under the global $Z_2$ spin reflections or inversions.

The chirality mechanism of Refs.~\onlinecite{Kawamura92} and
\onlinecite{Kawamura2}  takes the following 
two-step strategy in explaining the 
real SG transition: The first step concerns with the property of the
fully isotropic Heisenberg SG,
an idealization of experimental SG materials. The chirality scenario 
claims that the fully isotropic Heisenberg SG exhibits a finite-temperature CG
transition without the conventional SG long-range order. The CG
transition breaks only the $Z_2$ spin-reflection symmetry with keeping
the $SO(3)$ spin-rotational symmetry. The occurrence of the CG transition
necessarily entails the spin-chirality decoupling. 

Obviously, such a scenario does not apply to 
the infinite dimensional limit, {\it i.e.\/}, to the
mean-field Heisenberg Sherrington-Kirkpatrick (SK) model, in which 
the spin itself, not the chirality, behaves as an
order parameter of the transition. 
Due to the noncoplanar nature of the spin configuration in the SG state, 
the SG long-range order
trivially accompanies the CG long-rage order, whereas the opposite is 
not necessarily true.
One should note that, in the conventional case where
the spin variable is a proper
order parameter of the transition as in the case of the SK model, 
the chirality, which is given by the multiple of the spin, 
exhibits a less singular behavior than the spin at the SG transition.
In fact, the chirality shows only moderate behavior at the SG
transition of the mean-field Heisenberg SK
model in which the spin, not the chirality, is the order parameter of the 
transition.\cite{ImagawaKawamura03}

In contrast to the mean-field model or the high-dimensional Heisenberg SG
models, the
problem could be very non-trivial in lower dimensions where the
order-parameter fluctuation might change the nature of ordering dramatically.
At present, there seems to be no consensus about the lower critical dimension
$d_{\rm SG}^{\textrm{LCD}}$ of the SG order, while the corresponding
upper critical dimension is expected to be six. The CG order, if exists,
may emerge slightly above, at, or below $d_{\rm SG}^{\textrm{LCD}}$. 

It has been proved that the SG long-range order does not exist at any finite
temperature in the  two dimensional Heisenberg SG.\cite{Schwartz}
The numerical
domain-wall renormalization-group calculation as well as the MC
simulation suggested that both the spin
and the chirality ordered only at zero temperature\cite{Kawamura92}. 
Interestingly, however, the estimated SG and CG 
correlation-length exponents at this $T=0$ transition 
differ significantly from each other, {\it i.e.\/},
$\nu_{\rm CG} > \nu_{\rm SG}$.\cite{KawaYone}
This implies that in 2D 
the spin and the chirality are decoupled at long length
scale, the chirality dominating 
the long-length behavior. 

In view of these transition behaviors of the Heisenberg SGs, 
it appears likely that 
the principal player in long-scale phenomena
changes from the spin to the chirality as the spacial dimensionality is
decreased. Thus, the behavior in dimension three  is  
the current issue, which is the subject of
the present work. 

The second step of the chirality mechanism concerns with
the effect of the random magnetic anisotropy which inevitably exists
in real SG  magnets. 
The random anisotropy energetically 
breaks the $SO(3)$ spin-rotation symmetry in the
Hamiltonian, retaining the $Z_2$ inversion  symmetry only. 
When the anisotropic system exhibits the CG long-range order with 
spontaneously breaking the $Z_2$ 
inversion symmetry, there no longer remains any global symmetry
degree of freedom to leave the system in the paramagnetic phase. 
Hence, once the CG order occurs in the presence of the random anisotropy, 
the spin degree of freedom also behaves like the chirality. 
This is the spin-chirality recoupling due to the random magnetic anisotropy.

Such an anomaly revealed in the spin sector
via the random magnetic anisotropy
can be detected experimentally by standard magnetic measurements
\textit{e.g.\ }, as a divergence of the nonlinear susceptibility
\textit{etc\ }, 
whereas the CG
long-range order is difficult  to observe experimentally. 

We note that in this mechanism the anisotropy plays only 
a secondary role: 
The anisotropy certainly
reduces the symmetry of the Hamiltonian relative to the fully isotropic system,
but does {\it not change the broken symmetry of the transition\/}. 
The critical properties of the CG transition and of the 
low-temperature CG phase
are expected to be not affected by the magnetic anisotropy,  which, however, 
are now directly
observable via the standard spin-related quantities. 
This chirality scenario predicts that experimentally observed SG
transitions belong to the same universality class as that of the
the CG transition
of the fully isotropic model. It is thus highly interesting to clarify the
nature of the phase transition of the ideal isotropic Heisenberg SG.

\subsection{Spin-chirality decoupling/coupling scenario}
\label{sec:decoupling}

\begin{figure}
\includegraphics[scale=0.65]{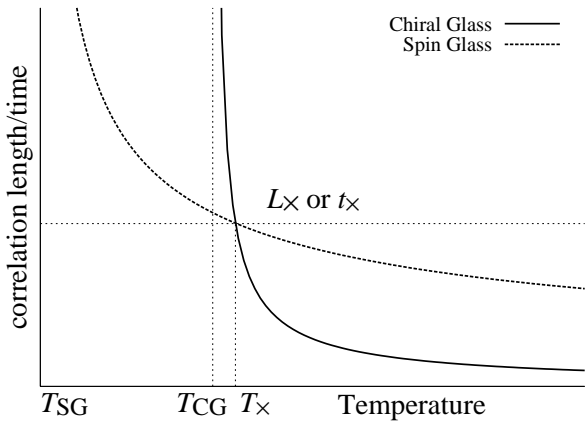}
\caption{A schematic figure of the crossover between the spin-glass and
 chiral-glass correlation lengths (correlation times) expected in the chirality
 mechanism.
According to the chirality mechanism, the CG correlation length (correlation time) 
diverges
 toward the CG transition  temperature $T=T_{\rm CG}$, while the SG one
 diverges at a lower  temperature $T=T_{\rm SG} < T_{\rm CG}$.  
}
\label{fig:cross-xi}
\end{figure}

As mentioned above, in lower dimensions, a relevant
degree of freedom which dominates 
the long-scale phenomena might well change from the spin to the chirality. 
The chirality scenario expects
that in 3D  there exists
a crossover temperature $T_{\times}$ which separates the two
temperature regimes,  as illustrated in Fig.~\ref{fig:cross-xi}. 
In  the higher temperature regime, 
the SG correlation length is longer than the CG correlation length,
dominating the long-scale phenomena. This is simply due to the fact that
the sensible definition of the local chirality requires the development
of the spin short-range order of at least a few lattice spacings.
As the temperature is decreased, both the SG and CG correlation lengths
grow, but at  different rates, so that the CG correlation length
eventually outgrows the SG correlation length at the crossover
temperature $T_{\times}$. An example of such a  crossover behavior
between the spin and the chiral correlations can be seen 
explicitly in a certain toy model: See Fig. 10 of Ref.~\onlinecite{UdaYoshino}.
Then, the relevant degree of freedom for the long-scale
behavior changes at $T_{\times}$ from the spin to the chirality.  
Below $T_{\times}$, the long-scale phenomena are governed by the CG
correlation, not by the SG one. This is the spin-chirality decoupling 
expected to occur in the fully isotropic model. 

Let us discuss in some detail the  finite-size effect inherent to
the simulation data in the critical region. 
The situation here is not simple because the system has two length scales,
each associated with the spin and with the chirality. 
Suppose that the CG transition occurs at $T=T_{\rm CG}$ without the
conventional SG long-range order. 
Then, the crossover temperature $T_{\times}$ at
which the CG correlation length outgrows the SG correlation length should
be located somewhat above $T_{\rm CG}$: See Fig.1.  A necessary condition
 for detecting
the spin-chirality decoupling is that the measurement temperature 
lies below $T_{\times}$. It is, however, not enough.   
At a temperature below $T_{\times}$, 
one needs to probe the system beyond the
crossover length above which the spin-chirality decoupling becomes eminent.
Thus,  a large-size simulation exceeding the crossover length is required in order
to detect the spin-chirality decoupling. 
Unfortunately, the crossover length scale is unknown {\it a priori\/}, 
and is to be
investigated by numerical simulations.  
A natural criterion might be that it is given by the SG correlation
length at the crossover temperature $T_{\times}$ as shown in
Fig.~\ref{fig:cross-xi}. 
Even in the CG  ordered phase, 
the spin-chirality decoupling might hardly be observable at the length scale
below the crossover length. Rather, 
it is natural to expect that the trivial spin-chirality coupling is observed
below the crossover length scale because the chirality is a composite
operator of the spin on the short scale of lattice spacing not independent
of the spin, roughly being $\chi \approx S^3$.

In the  CG critical region, the chirality-related
quantities should exhibit the true asymptotic 
critical behavior, {\it e.g.\/},  
a power-law singularity characterized by the associated CG exponents.
At short length scales below the
crossover length scale, due to the trivial coupling between the spin and 
the chirality,
even the spin-related quantities are expected to
exhibit the similar critical behavior to the chirality-related
quantities. Namely, up to the crossover length scale, 
it seems as if the SG order developed as a long-range order. 
It is intrinsically 
difficult at shorter length scales 
to distinguish such a pseudo-critical behavior induced by
the CG long-range order from 
the true SG long-range order.
Hence, it is crucially important to
estimate the crossover length scale and to study the long-scale behavior of the
system beyond this length scale.

\begin{figure}
\includegraphics[width=\figwidth]{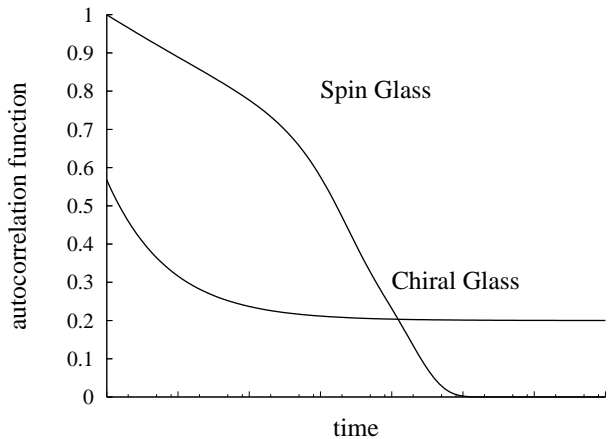}
\caption{A schematic figure of the time evolution of the spin-glass and
 chiral-glass two-time autocorrelation functions.}
\label{fig:cross-c}
\end{figure}

Essentially the same argument also applies to the case of the temporal scale. 
As an example, we discuss here the behavior of the autocorrelation 
functions
based on the notion of the crossover time scale. 
Figure~\ref{fig:cross-c}  shows the schematic representation of the behavior
of the SG and CG autocorrelation functions below $T_{{\rm CG}}$
expected  from the spin-chirality coupling/decoupling picture. 
In the long-time limit, 
the chirality autocorrelation function is expected
to saturate to a certain finite value
after an initial fast decay, while the spin autocorrelation
function is expected to decay toward zero
asymptotically. A comment is in order concerning the 
transient behavior of the spin 
autocorrelation function at time scales shorter than the crossover time
scale:  At shorter time scales, 
the spin autocorrelation function
might exhibit the pseudo-ordering feature 
dictated by the CG one through
the trivial spin-chirality coupling. 
This might well lead to a hump-like pseudo-ordering structure in the
time dependence of the spin autocorrelation function as shown in Fig.2, 
which, however, does not persist in the
long-time limit beyond the crossover time scale. 
This means that, in order to properly discuss the true asymptotic behavior 
of the dynamics of the model,
a particular attention should be paid to the crossover time
scale.

\section{The Model and the Monte Carlo method}
\label{sec:model}

We study a classical Heisenberg model defined
by the Hamiltonian,  
\begin{equation}
 H(\vec{S})=-\sum_{\langle ij\rangle}J_{ij}\vec{S}_i\cdot\vec{S}_j, 
\label{eqn:model}
\end{equation}
where $\vec{S}_i=(S_i^x,S_i^y,S_i^z)$ is a three-component unit vector,
and the summation runs over all nearest-neighbor pairs.
The lattice is a simple-cubic lattice with the total number
of $N=L^3$ sites. The nearest-neighbor couplings $J_{ij}$ take the values
$\pm J$ randomly with equal probability.  
Periodic boundary conditions are imposed for all the directions. 
The lattice sizes studied are $L=8$, $12$, $16$ and $20$, where the sample
average is taken over 976 ($L=8$), 964 ($L=12$), 280 ($L=16$), and 32 ($L=20$)
independent bond realizations.  

We perform an equilibrium MC simulation of the model. 
In our simulation, we make use of the exchange MC method,\cite{EMC}
which is also called parallel tempering.\cite{Marinari96} 
In the exchange MC method, one MC step consists of two elementary
updates, a standard single-spin heat-bath flip\cite{Olive86} and an
exchange trial of spin configurations at neighboring temperatures. 
The latter reduces the slow relaxation at low temperatures with the help of
the high-temperature fast dynamics. 
The method has turned out to be quit efficient in thermalizing  a wide
class of hardly-relaxing systems such as SG systems and proteins. 
We ensure equilibration by checking that various observables attain 
stable values, 
no longer 
changing with the amount of MC steps: See 
Refs.~\onlinecite{ImagawaKawamura03} and \onlinecite{KawaHuku99} for
further details of the equilibration procedure.
Our MC simulations have been performed up to the size $L=20$ and 
up to the temperature $T/J=0.15$.
This could be achieved only by using the exchange MC method. 
The numbers of temperature points used in our exchange MC method are 32
for $L=8, 12, 16$, and $48$ for $L=20$. 

Error bars are estimated via sample-to-sample fluctuations for
the linear quantities such as the
order parameters, and by the jackknife method for the non-linear quantities
such as the Binder parameter and the correlation length mentioned below.

\section{Physical Observables}
\label{sec:pq}

In the present section, we introduce various physical
quantities observed in our simulations, and discuss some of their basic 
properties.  

In glassy systems, it is often convenient to
define  as an order parameter an overlap variable between two 
independent systems with the same
Hamiltonian. For the Heisenberg spin,
the  overlap may be defined as a tensor variable
between the $\mu$ and $\nu$ components ($\mu, \nu =x,y,z$) 
of the Heisenberg spin by
\begin{equation}
 q_{\mu\nu}=\frac{1}{N}\sum_{i=1}^{N} S_{i\mu}^{(1)}S_{i\nu}^{(2)}, 
\end{equation}
where the upper suffixes (1) and (2) denote the two replicas of the system with
the same interaction set. 

A chiral overlap is defined in terms of the local chiral variable 
(\ref{scalar-chirality}) by 

\begin{equation}
 q_\chi=\frac{1}{3N}\sum_{i\mu}\chi_{i\mu}^{(1)}\chi_{i\mu}^{(2)}. 
\end{equation}

The squared SG order parameter is then given by 
\begin{equation}
q_{\rm SG}^{(2)}=\left[\left\langle\sum_{\mu\nu}q_{\mu\nu}^2\right\rangle\right],  
\label{eqn:q2sg}
\end{equation}
where $\langle\cdots\rangle$ denotes a thermal average and $[\cdots]$
denotes an average over the bond disorder. 
The corresponding squared CG order parameter is defined by 
\begin{equation}
q_{\rm CG}^{(2)}=\frac{\left[\left\langle q_{\chi}^2\right\rangle\right]}{\overline{\chi}^4}, 
\label{eqn:q2cg}
\end{equation}
which is normalized by the mean-square amplitude  of the local
chirality, 
 \begin{equation}
  \overline{\chi}^2 = \frac{1}{3N} \sum_i^N \sum_\mu [\langle\chi_{i\mu}^2\rangle]. 
 \end{equation}
The local chirality amplitude remains non-zero only when the 
spins have a non-coplanar structure locally. This quantity
weakly depends on the temperature, in contrast to the Heisenberg spin variable 
whose amplitude is  fixed to be unity by definition. 
In the high-temperature symmetric phase, these SG and CG
order parameters are
essentially equivalent to the associated SG and CG
susceptibilities defined by
$\chi_{\rm SG}=Nq_{\rm SG}^{(2)}$ and $\chi_{\rm CG}=3Nq_{\rm
CG}^{(2)}$, respectively.

A standard finite-size scaling of the second-order transition for the
equilibrium SG and CG order parameters takes the form
\begin{equation}
 q^{(2)}\sim L^{-(1+\eta)}f\left(|T-T_{\rm c}|L^{1/\nu}\right),
\label{eqn:fss-c}
\end{equation}
where $\nu$ is
the exponent of the correlation length,
and $\eta$ is the exponent describing the decay of the 
correlation function at the critical point $T=T_{\rm c}$.
At $T_{\rm c}$, the order parameter decays as a power law 
with the size $L$,
\begin{equation}
 q^{(2)}\propto L^{-(1+\eta)}.
\label{eqn:fss-c-l}
\end{equation}

One often uses the  Binder parameter to estimate the critical
temperature. In the Heisenberg SG, the Binder parameter for the
SG order is defined by
\begin{equation}
 g_{\rm SG}=\frac{1}{2}\left(11-9
\frac{[\langle q^4\rangle]}
{[\langle q^2\rangle]^2}\right), \ \ \ \ \
q^2 = \sum_{\mu, \nu} q_{\mu \nu}^2,  
\label{eqn:bin-q}
\end{equation}
while that for the CG oder is defined by 
\begin{equation}
 g_{\rm CG}=\frac{1}{2}\left(3-\frac{[\langle q_\chi^4\rangle]}{[\langle q_\chi^2\rangle]^2}\right). 
\label{eqn:bin-c}
\end{equation}
In the thermodynamic limit, these Binder parameters
are normalized to unity  in the
non-degenerate ordered state, and to zero in the high-temperature disordered
state.  Since the Binder parameter is a dimensionless quantity,
and the dimensionless quantity should be size-independent at the
critical temperature $T_c$, the Binder parameters of different system sizes
plotted as a function of temperature should yield 
a crossing or merging point at $T_c$.

In terms of the $k$-dependent overlap variable, one can define the 
Fourier-transformed two-point CG and SG correlation functions.
For the CG, the $k$-dependent chiral-overlap is defined by
\begin{equation}
q_\chi(\vec k)=\frac{1}{N}\sum_{i=1}^N\chi_{ix}^{(1)}\chi_{ix}^{(2)} 
\exp (i\vec k\cdot \vec r_i), 
\end{equation}
in which the chiral variable along the $x$-axis, $\chi_{ix}$, is considered.
The Fourier-transformed CG correlation function is then defined by
\begin{equation}
q^{(2)}_\mathrm{CG}(\vec{k})=\left[\left\langle|q_\chi(\vec k)|^2\right\rangle\right].
\end{equation}

For the SG, the $k$-dependent spin-overlap is defined by
\begin{equation}
q_{\mu \nu}(\vec k)=\frac{1}{N}\sum_{i=1}^N 
S_{i\mu}^{(1)}S_{i\nu}^{(2)}
 \exp (i\vec k\cdot \vec r_i), 
\end{equation}
whereas the Fourier-transformed SG correlation function is defined by
\begin{equation}
q^{(2)}_{\rm SG}(\vec{k})=\left[\left\langle\sum _{\mu \nu}|q_{\mu \nu}(\vec k)|^2\right\rangle\right]. 
\end{equation}
Via these CG and SG Fourier-transformed correlation functions, the associated CG and SG  
finite-system correlation lengths are defined by
\begin{equation}
\xi = 
\frac{1}{2\sin(k_\mathrm{m}/2)}
\sqrt{\frac{q^{(2)}(\vec 0)}
{q^{(2)}(\vec{k}_\mathrm{m})}-1}
\ \ ,
\end{equation}
%
%
where $\vec{k}_{\rm m}=(2\pi/L,0,0)$ and $k_{\textrm{m}}=|\vec k_{\textrm{m}}|$. 

One can then define a dimensionless quantity,
the normalized correlation length $\xi_{{\rm CG}}/L$ and $\xi_{{\rm SG}}/L$. 
Since $\xi/L$ is dimensionless,
it should exhibit the same scaling 
property as the Binder parameter. 
Thus, the ratio $\xi/L$ for different $L$ should cross or merge at the
critical temperature.

In probing the nature of the low-temperature glassy ordered phase, one
useful quantity is the distribution of the 
overlap.  The chiral-overlap distribution is defined by 
\begin{equation}
 P(q'_\chi)=\left[\left\langle\delta(q'_\chi-q_\chi)\right\rangle\right]. 
\label{eqn:dist-c}
\end{equation}
The squared CG order parameter $q_{\rm CG}^{(2)}$ defined above
is the second moment of the chiral-overlap distribution function. 

The spin-overlap distribution is defined originally in the tensor
space with $3\times 3=9$ components. To make the quantity more easily tractable,
one may define the diagonal spin overlap which is  the trace of the original 
tensor overlap, and introduce the associated diagonal-spin-overlap 
distribution by
\begin{equation}
 P(q_{\rm diag})=\left[\left\langle\delta\left(q_{\rm diag}-\left(\sum_{\mu=x,y,z}q_{\mu\mu}\right)\right)\right\rangle\right].   
\label{eqn:dist-q}
\end{equation}
This distribution function is symmetric with respect to $q_{\rm diag}=0$,
and is expected to be a Gaussian distribution around $q_{\rm diag}=0$ in the
high-temperature disordered phase. 
Reflecting the fact that 
the diagonal-spin-overlap is not invariant under the global $O(3)$ spin
rotation, $P(q_{\rm diag})$ in the possible SG ordered phase
develops a nontrivial shape, not just consisting of
the delta-function peaks related to
$q_{\rm EA}$, even when the ordered state is a trivial one simply
described by a self-overlap $q_{\rm EA}$.\cite{ImagawaKawamura03}
If the possible SG ordered state 
accompanies a replica-symmetry breaking (RSB), 
further nontrivial structures would be added to 
$P(q_{\rm diag})$. Meanwhile,
it is recently shown in Ref.~\onlinecite{ImagawaKawamura03}  that, 
in the possible SG ordered state, the diverging peaks
corresponding to the self-overlap necessarily appears in $P(q_{\rm diag})$ 
at $q_{\rm diag}=\pm \frac{1}{3}q_{\rm EA}$ in the thermodynamic limit. 
Hence, the existence or nonexistence of these divergent peaks 
could be used as an unambiguous measure of
the possible SG long-range order in the Heisenberg SG, 
irrespective the occurrence of the RSB.  

Another interesting feature of glassy systems might be the nature of 
their 
sample-to-sample fluctuations, particularly its possible non-self-averageness.
As an indicator of the lack of self-averageness, one may use the so-called $A$
parameter.\cite{Marinari99} 
For the CG order, it is  defined by 
\begin{equation}
A_{\rm CG}(T)\equiv \frac{[\langle q_{\chi}^2\rangle^2]-[\langle q_{\chi}^2\rangle]^2}{[\langle q_{\chi}^2\rangle]^2},
\end{equation}
while for the SG order, 
\begin{equation}
A_{\rm SG}(T)\equiv \frac{[\langle q^2\rangle^2]-[\langle q^2\rangle]^2}
{[\langle q^2\rangle]^2}.
\end{equation}
The order parameter is 
non-self-averaging when the associated
$A$ parameter is non-zero, and is self-averaging when $A$ is equal to zero.

One can also define the so-called Guerra parameter
$G$.\cite{Guerra}
For the CG order, it is defined by
\begin{equation}
G_{\rm CG}(T)\equiv \frac{[\langle q_{\chi}^2\rangle^2]-
[\langle q_{\chi}^2\rangle]^2}
{[\langle q_{\chi}^4\rangle]-[\langle q_{\chi}^2\rangle]^2},
\end{equation}
while for the SG order,
\begin{equation}
G_{\rm SG}(T)\equiv \frac{[\langle q^2\rangle^2]-[\langle q^2\rangle]^2}
{[\langle q^4\rangle]-[\langle q^2\rangle]^2}.
\end{equation}
Unlike the $A$ parameter, the $G$ parameter can take a non-zero value 
even when the ordered state is a trivial one without accompanying the 
RSB.\cite{Bokil,Marinari98} 
The $G$ parameters are related to the $A$ parameters and the Binder 
parameters $g$ via the relations,
\begin{equation}
A_{{\rm CG}}=2(1-g_{{\rm CG}})G_{{\rm CG}},
\end{equation}
\begin{equation}
A_{{\rm SG}}=\frac{2}{9}(1-g_{{\rm SG}})G_{{\rm SG}}.
\label{eqn:GAg}
\end{equation}
These relations indicate that, so long as the Binder parameter $g$
takes any value 
different from unity in the ordered phase, a non-zero $A$ necessarily means 
a non-zero $G$. By contrast, if the Binder parameter $g$ takes a value unity 
in the ordered phase, a non-zero $A$ may or may not mean a non-zero $G$.

Information about the equilibrium dynamics can be obtained 
from the spin and chiral two-time autocorrelation functions defined by 
\begin{eqnarray}
 C_s(t) & = &\frac{1}{N}\sum_i[\langle \vec S_i(t_0)\cdot 
\vec S_i(t+t_0)\rangle], \\
 C_\chi(t) & = &\frac{1}{3N}\sum_{i\mu}[\langle \chi_{i\mu}(t_0)\chi_{i\mu}(t+t_0)\rangle], 
\label{eqn:autocorrelation}
\end{eqnarray}
where the time evolution in our MC simulation 
is made according to the standard heat-bath
updating not accompanying the exchange process. Initial spin
configurations at $t=t_0$ are taken from 
equilibrium spin configurations generated in our exchange MC runs.
Below the transition temperature $T_c$ (if any), 
these autocorrelation functions
converge in the long time limit to the Edwards-Anderson SG and CG  
order parameters, whereas above $T_c$ these
autocorrelation functions decay exponentially toward zero with a characteristic
correlation time,  which
diverges as the temperature $T$ approaches $T_c$.
Just at $T_c$, the autocorrelation functions exhibit a power-law decay, 
\begin{equation}
 C(t) \sim t^{-\beta/z\nu}, 
\label{eqn:qoft-tc}
\end{equation} 
where $z$ is the dynamical critical exponent. 
These features are described by the standard bulk dynamical scaling form, 
\begin{equation}
C(t) \sim |T-T_{\rm c}|^\beta f(t |T-T_{\rm c}|^{z\nu}), 
 \label{eqn:dss}
\end{equation} 
where $f(x)$ is a scaling function whose asymptotic forms for $x\ll 1$
and $x\gg 1$ are $x^{-\beta/z\nu}$ and $\exp(-x)$, respectively. 

\section{The Numerical results}
\label{sec:NR}

In the present section, we show our MC results of the three-dimensional
isotropic Heisenberg SG model.

\subsection{Binder parameter}
\label{subsec:BP}

\begin{figure}
\includegraphics[width=\figwidth]{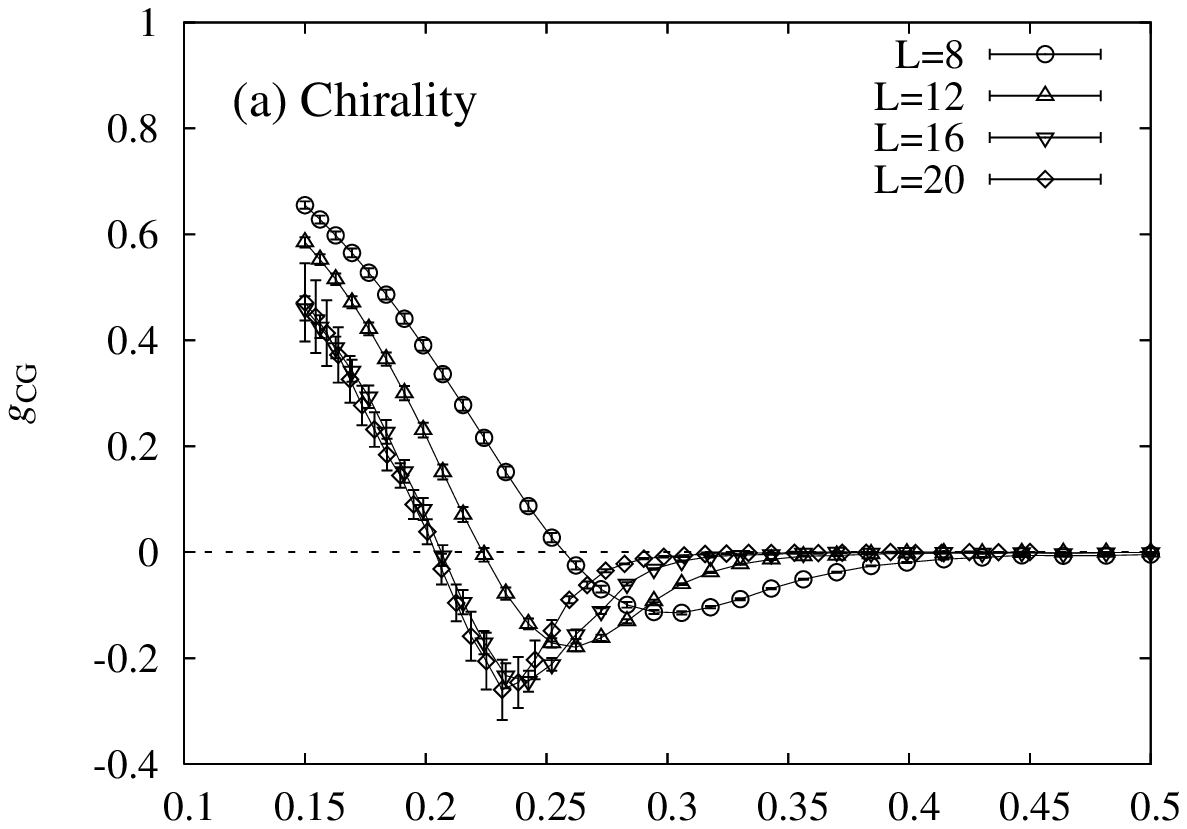}
\includegraphics[width=\figwidth]{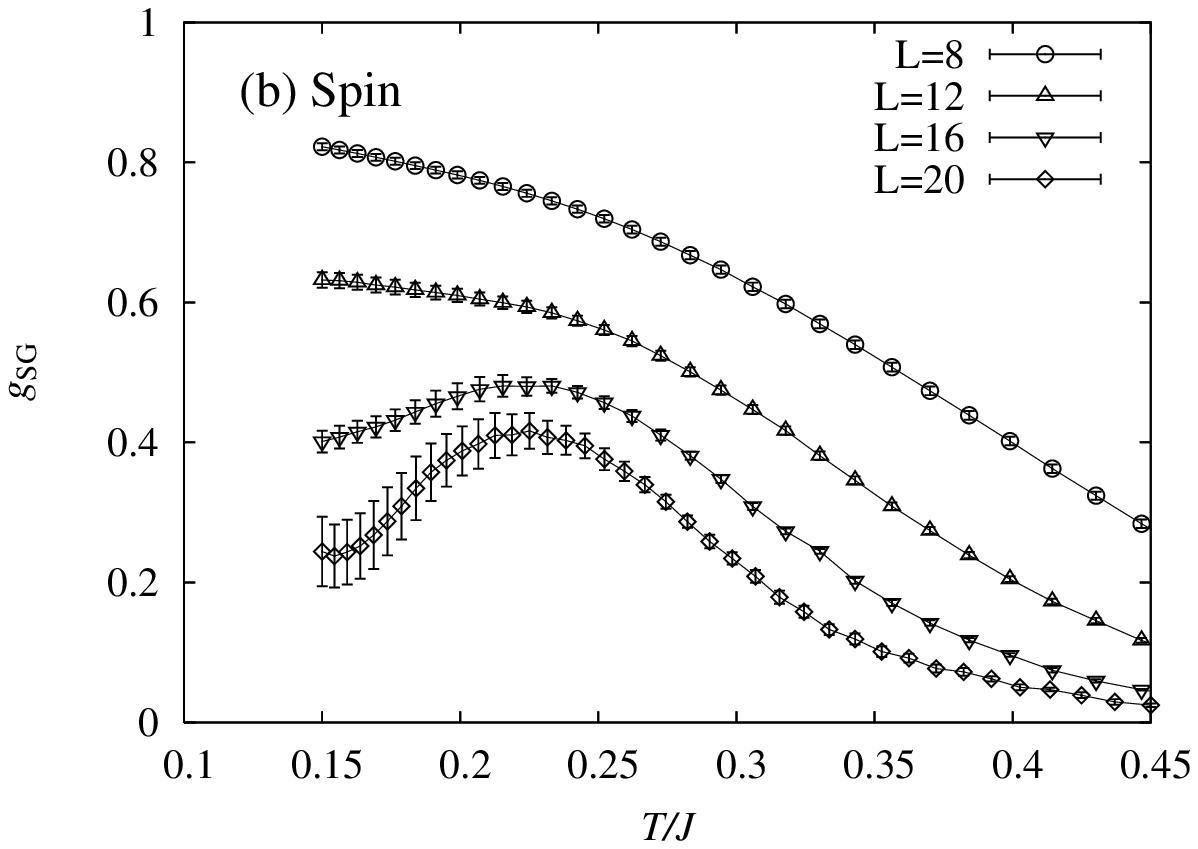}
\caption{
The temperature and size 
dependence of the chiral-glass Binder parameter; upper figure (a),
and of the spin-glass Binder parameter; lower figure (b), 
of the 3D $\pm J$ Heisenberg SG.   
}
\label{fig:bin-d00}
\end{figure}

\begin{figure}
\includegraphics[width=\figwidth]{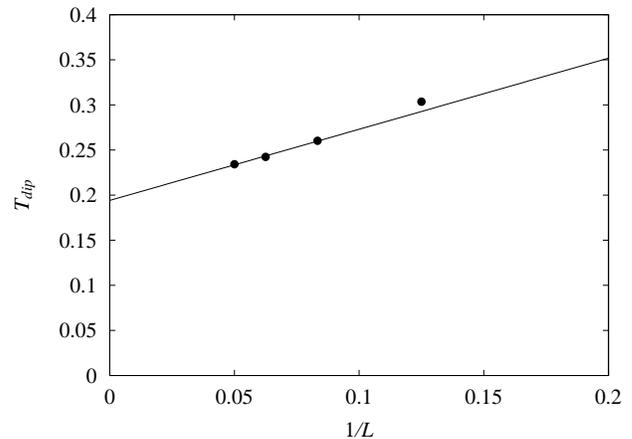}
\caption{
The dip temperature of the chiral-glass Binder parameter
$g_{\rm CG}$ is plotted against $1/L$. 
The solid line represents a linear fit of the data. 
Its extrapolation to the $L\rightarrow \infty$ limit gives
an estimate of the bulk
chiral-glass transition temperature, $T_{\rm CG}/J=0.194(5)$. 
}
\label{fig:dip-c00}
\end{figure}

In Fig.~\ref{fig:bin-d00}, we show the temperature and size dependence of 
the Binder parameters  
both for the chirality; upper figure (a), and for
the spin; lower figure (b). As can be seen from Fig.~\ref{fig:bin-d00}(a),
a crossing of the CG Binder parameter 
$g_{\rm CG}$ of different $L$
is observed on the negative side of $g_{\rm CG}$, 
{\it not\/} on the positive side as in the standard cases. 
With increasing $L$, the crossing temperature gradually shifts toward
lower temperatures. 
A behavior similar to this 
has also been observed in the Binder parameter of other
models, including the Heisenberg SG\cite{HukuKawa00a,ImagawaKawamura03}
and the mean-field SG\cite{HukuKawa00b,Picco01}. In particular,
in a class of mean-field SG models exhibiting a one-step RSB, 
the Binder parameter at the transition point
$T_{\rm c}$ takes a negative
value, sometimes even negatively divergent.\cite{HukuKawa00b,Picco01}
It implies that the temperature at which the Binder parameter  
for finite $L$ takes a minimum,  
a dip temperature $T_{\rm dip}(L)$, 
approaches the critical temperature, {\it i.e.\/}, 
$T_{\rm dip}(L)\rightarrow T_{\rm c}$ as $L\rightarrow \infty$. 
Recently, this
method of estimating the bulk transition temperature
was successfully applied to the Heisenberg 
SG.\cite{ImagawaKawamura02,ImagawaKawamura03}
In Fig.~\ref{fig:dip-c00}, we plot $T_{\rm dip}(L)$
against $1/L$. 
An extrapolation to the thermodynamic limit $1/L\rightarrow 0$
gives us an estimate of the bulk
CG transition temperature, $T_{\rm CG}/J=0.194(5)$.

By contrast, as shown in 
Fig.~\ref{fig:bin-d00}(b),  the SG
Binder parameter $g_{\rm SG}$ monotonically decreases toward zero 
with increasing $L$ at all
temperatures studied. There is no signature of the transition in the
investigated temperature range, 
no negative dip nor the crossing, in contrast to the CG
Binder parameter. 
This suggests that the SG transition temperature, if any, is
located at a temperature lower than the
temperature range studied here.  Fig.~\ref{fig:bin-d00}(b) reveals,
however, that 
an anomalous bend appears in $g_{\rm SG}$ for larger sizes $L\geq 16$ 
at around $T/J\simeq 0.22$, close to the CG
transition temperature, although $g_{\rm SG}$ never becomes
size-invariant at any temperature, 
as it should have been in a second-order transition. The reason
why $g_{\rm SG}$ exhibits such an anomalous bend around  $T_{\rm CG}$ 
might be understood as follows: 
At the CG transition, a reflection 
symmetry is spontaneously broken and the entire phase space is divided into 
ergodic components, in each of which a proper-rotational symmetry is 
still preserved. 
As a result, the ordering behavior of the Heisenberg
spin would change at $T_{\rm CG}$, 
though the spin itself does not order even below $T_{\rm CG}$.
We note that a similar bend in  $g_{\rm SG}$ has also been observed in the
{\it two-dimensional\/} Heisenberg SG \cite{KawaYone} where the absence of 
a finite-temperature SG transition has been well established.\cite{Schwartz}
 
It is sometimes argued in the literature
that the  Binder-parameter analysis might
not work in the SG problem. Such a suspicion might partly be 
based on the observation that only weak merging behavior was observed
at the SG transition temperature of the three-dimensional EA Ising
model which is believed to exhibit a finite-temperature  SG 
transition.\cite{BhattYoung,Ogielski,KawashimaYoung}
As long as the SG long-rage order really sets in 
at finite temperatures, however, 
it is hardly conceivable that the Binder parameter  
for asymptotically large lattices exhibits a 
non-singular behavior only. 
In particular, the Binder parameter 
should become scale-invariant at
the SG transition point, so long as the transition is continuous.
Indeed, in a recent MC simulation 
of the {\it mean-field\/} Heisenberg SG,\cite{ImagawaKawamura03}
which is known to exhibit a non-zero SG long-range order below $T_{\rm SG}$,
a clear crossing of the SG Binder parameter $g_{\rm SG}$  has been observed 
at $T_{\rm SG}$, in sharp contrast to
our present data of Fig.~\ref{fig:bin-d00}(b). 

Then, one might argue that the finite-size effect would be
significant here in $g_{\rm SG}$ and the large-$L$ asymptote might still 
be far away.
One sees from Fig.~\ref{fig:bin-d00}(a), however,  
that the CG Binder parameter $g_{\rm CG}$ for our two
largest sizes $L=16$ and 20 exhibits an almost
size-invariant behavior at and 
below $T_{{\rm CG}}$. If the Heisenberg spin orders
simultaneously with the chirality, and if the spin is the order
parameter of the transition and the chirality is only composite (secondary), 
it seems a bit hard to
understand why the chirality exhibits an almost scale-invariant near-critical
behavior for $L\geq 16$,
while the Heisenberg
spin still exhibits an off-critical scale-dependent behavior.
Hence, the behavior of $g_{\rm SG}$ observed in Fig.~\ref{fig:bin-d00}(b) 
remains
to be resolved if the occurrence of a finite-temperature SG transition 
is to be accepted in the investigated temperature range.

\subsection{Order parameter}
\label{sec:q2}

\begin{figure}
\includegraphics[width=\figwidth]{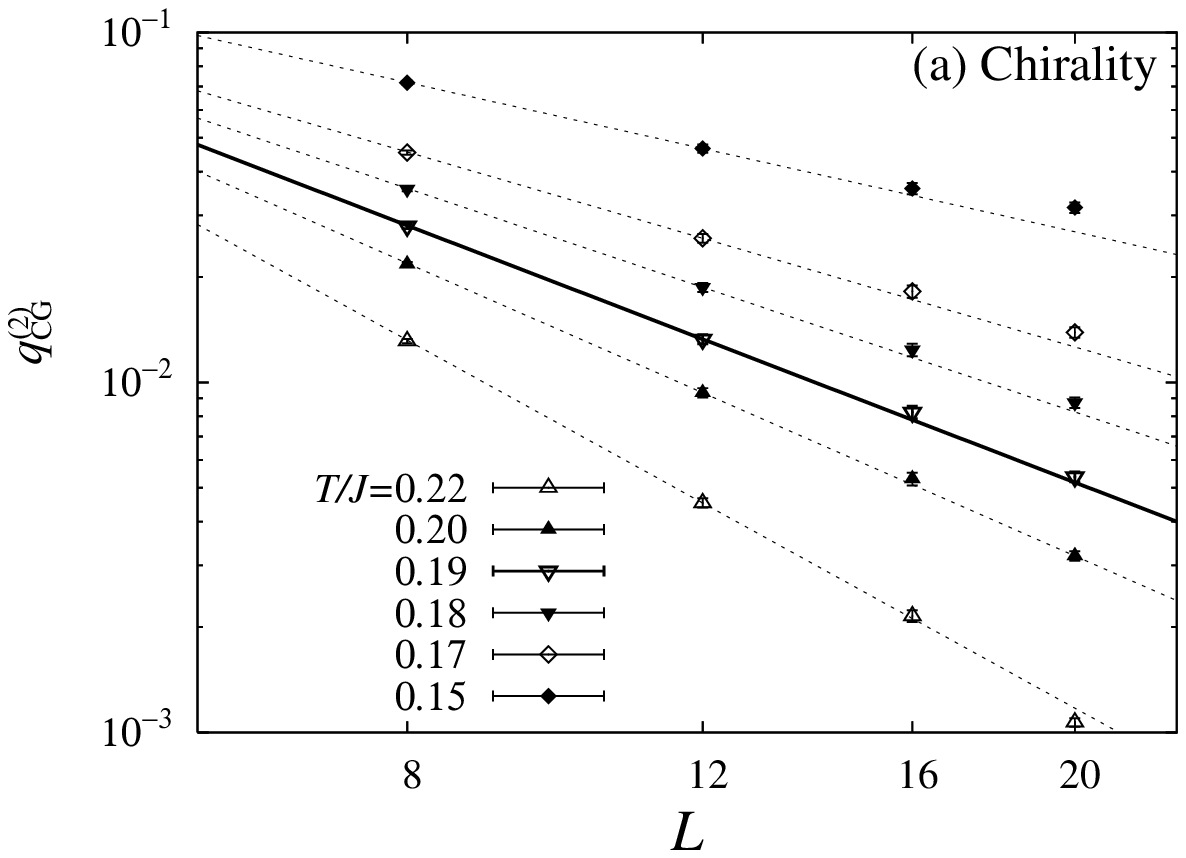}
\includegraphics[width=\figwidth]{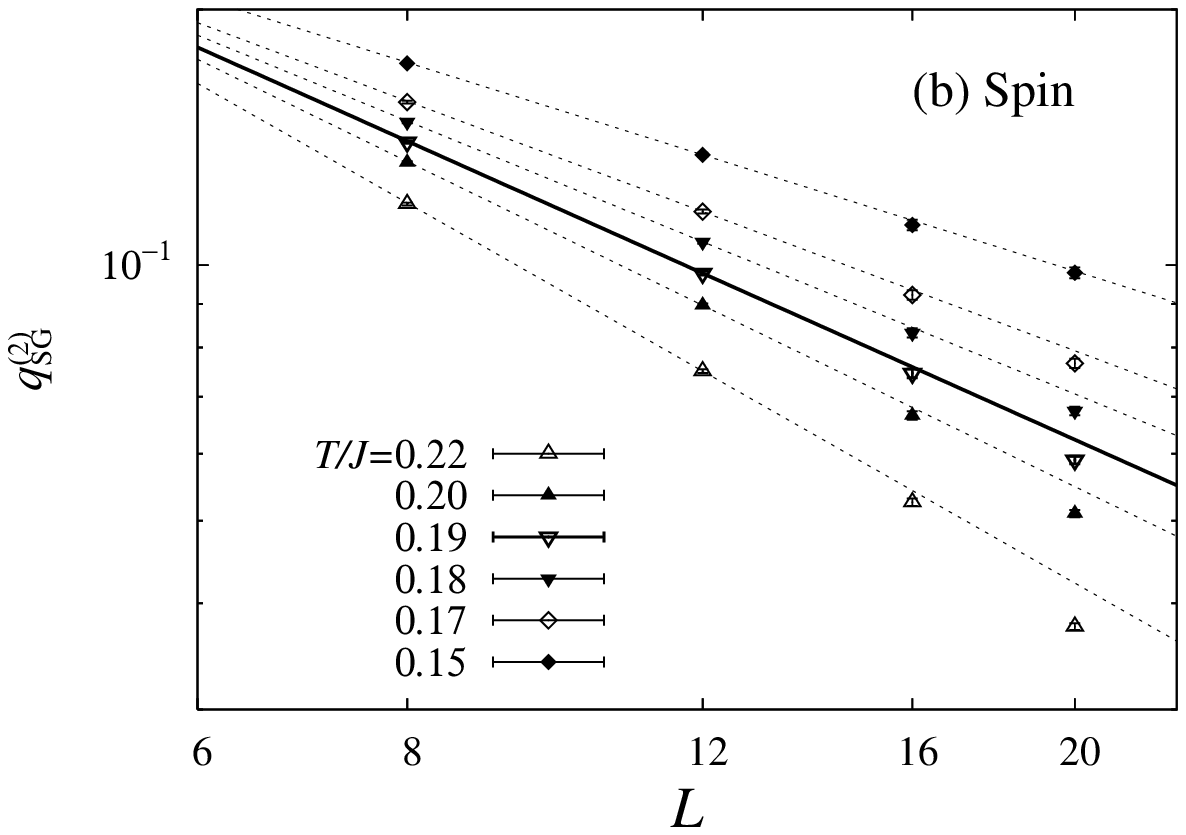}
\caption{
Double-logarithmic plot of the squared chiral-glass order parameter
$q_{\rm CG}^{(2)}$; upper figure (a), and of the squared
spin-glass order parameter $q_{\rm SG}^{(2)}$;  lower figure (b),
as a function of the system size $L$ for several
temperatures around the expected chiral-glass transition temperature.  
Straight lines are drawn by connecting the two data points of
$L=8$ and $12$ at each temperature. 
}
\label{fig:q2size}
\end{figure}

In Fig.~\ref{fig:q2size}, we show
the size dependence of 
the squared CG and SG order parameters, $q_{\rm CG}^{(2)}$ 
and $q_{\rm SG}^{(2)}$, 
for various temperatures.
In the upper figure (a), a double-logarithmic 
plot of the CG order  parameter $q_{\rm CG}^{(2)}$ is 
shown against the system size $L$.   
One generally expects that at $T_c$
the data of $q_{\rm CG}^{(2)}$ should lie on a straight line. 
In fact, the data of $q_{\rm CG}^{(2)}$
show a clear straight-line behavior around
$T/J=0.19$, which is close to the CG transition temperature
obtained by our analysis of $g_{\rm CG}$.  
The critical-point decay exponent
$\eta_{\rm CG}$ can be estimated from the slope of this straight line, 
yielding $1+\eta_{\rm CG}\sim 1.8$.
At higher temperatures, a deviation from the straight line, 
a downward trend, is observed indicative of the disordered phase.
At lower temperatures, particularly at our lowest temperature
simulated $T/J=0.15$,
the data of $q_{\rm CG}^{(2)}$ show a clear upward trend. 
This suggests  that this temperature is indeed 
below $T_{\rm CG}$, 
and that the low-temperature phase is a rigid one characterized by 
a non-zero $q_{\rm CG}^{(2)}$,
not likely to be
a critical phase like the Kosterlitz-Thouless (KT) phase.  

In  Fig.~\ref{fig:q2size}(b),  a double-logarithmic 
plot of the corresponding SG order parameter
$q_{\rm SG}^{(2)}$ is shown against the system size $L$.  
Again, the data are expected to lie on a straight line at the critical SG
transition temperature, if any.  
Such a straight-line behavior, however, is found only at our 
lowest temperature
studied $T/J=0.15$, whereas $q_{\rm SG}^{(2)}$ {\it never exhibits 
an upward trend  characteristic of the long-range ordered phase 
at any temperature studied\/}, in sharp contrast to the behavior 
of $q_{\rm CG}^{(2)}$. 
At temperatures higher than $T/J=0.15$, including the one at the
CG transition temperature $T/J\simeq 
0.19$, the data of $q_{\rm SG}^{(2)}$ show a linear behavior 
for smaller sizes, which
gradually changes into a downward trend for larger sizes. 
This can simply be interpreted as
a size-crossover which occurs around the length scale of
the SG correlation length at each temperature.  
We note that such a size-crossover
is clearly discernible even at a temperature $T/J=0.17$ 
which is below $T_{\rm CG}$. 
The length scale of the crossover, comparable to the spin correlation length,
grows as $T$ decreases, and  it is
considered to exceed our largest size $L=20$ at around $T/J=0.15$. 

This observation strongly suggests 
that the standard SG transition temperature of the model is lower than
$T/J=0.15$, and that, at least within the temperature range 
$0.15\alt T\alt 0.19$, 
solely the CG long-range order exists without the 
standard SG long-range order, {\it i.e.\/}, 
one has $T_{\rm CG}>T_{\rm SG}$.  

\begin{figure}
\includegraphics[width=\figwidth]{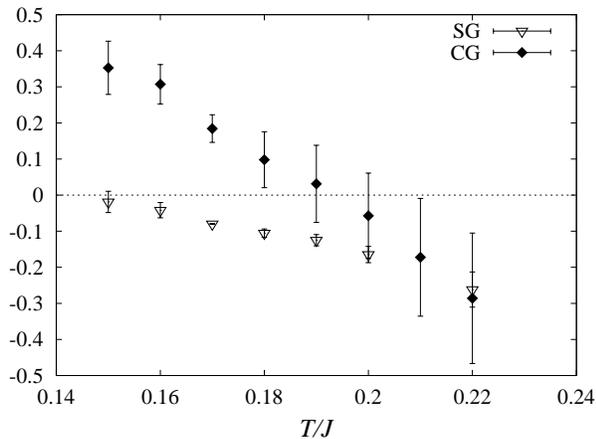}
\caption{
The curvatures of the $L$-dependence of the 
squared spin-glass 
order parameter $q_{\rm SG}^{(2)}$  (open symbols), 
and of the squared chiral-glass order parameter  
$q_{\rm CG}^{(2)}$ (filled symbols), 
are plotted versus the temperature.  
The curvature is expected to be zero at
the  respective transition temperature. 
} 
\label{fig:curve-d00}
\end{figure}

To make the situation more pronounced, 
we estimate following 
Refs.~\onlinecite{Katzgraber} and \onlinecite{KC}
the curvatures of the $L$-dependence 
of the two order parameters, $q_{\rm SG}^{(2)}$ and $q_{\rm CG}^{(2)}$, via 
second-order polynomial fits to the data of Figs.~\ref{fig:q2size}.
The curvature is expected to be zero at
the respective transition temperature. 
As shown in Fig.~\ref{fig:curve-d00},   the curvature for the CG
crosses the zero-axis around $T/J\simeq 0.19$, while that for the SG does
not cross the zero-axis there, but marginally touches on it at a 
lower temperature, $T/J\simeq 0.15$. 
The result indicates that the two transition temperatures, 
$T_{\rm SG}$ and
$T_{\rm CG}$, are well separated.

\subsection{Finite-size scaling of the order parameter}
\label{subsec:fss-q2}

\begin{figure}
\includegraphics[width=\figwidth]{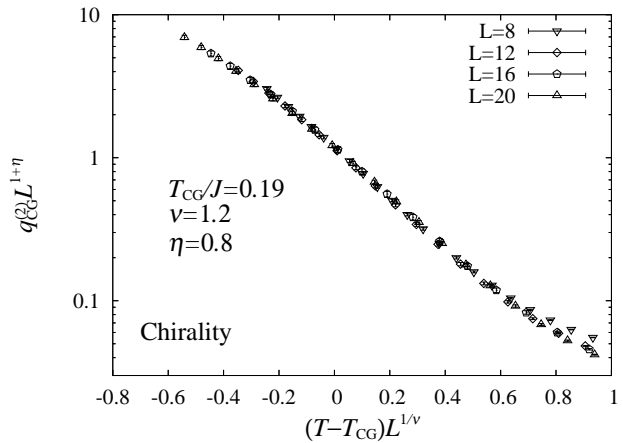}
\caption{
Finite-size-scaling plot of the squared chiral-glass order parameter. 
The best scaling is obtained with $T_{\rm CG}/J=0.19$, $\nu=1.2$ and
  $1+\eta=1.8$.} 
\label{fig:fss-c-d00}
\end{figure}

In order to estimate the correlation-length exponent $\nu$ associated
with the CG transition, we apply 
the standard finite-size scaling analysis to the squared CG order parameter
$q_{\rm CG}^{(2)}$ based on Eq.~(\ref{eqn:fss-c}). 
By taking $|T-T_{\rm CG}|/T_{\rm CG}L^{1/\nu}$ as the scaling variable, 
the best data collapse  is obtained with $T_{\rm CG}/J=0.19$, $\nu_{\rm
CG}=1.2$ and $\eta_{\rm CG}=0.8$.  
As shown in Fig.~\ref{fig:fss-c-d00}, the data both below and above
$T_{\rm CG}$ scale fairly well. 
If $|1/T-1/T_{\rm CG}|T_{\rm CG}L^{1/\nu}$ is taken as the scaling variable,
on the other hand, a slightly larger value of $\nu$, {\it i.e.\/}, 
$\nu_{\rm CG}=1.4$ and $\eta_{\rm CG}=0.7$, is preferred. The observed 
difference in the best values of the exponents might be 
due to the correction to scaling. Thus, we finally quote
$T_{\rm CG}/J=0.19(1)$, $\nu_{\rm CG}=1.3(2)$ and $\eta_{\rm CG}=0.8(2)$.
The error bar is estimated  by examining the quality of the fits with
varying the scaling parameters.
The estimated values of critical exponents are compatible with the
previous values obtained for the Heisenberg SG model 
but with the Gaussian bond distribution,\cite{HukuKawa00a} and also with
those for the $\pm J$ Heisenberg SG under external
fields \cite{KawamuraImagawa01,ImagawaKawamura02}. 

\begin{figure}
\includegraphics[width=\figwidth]{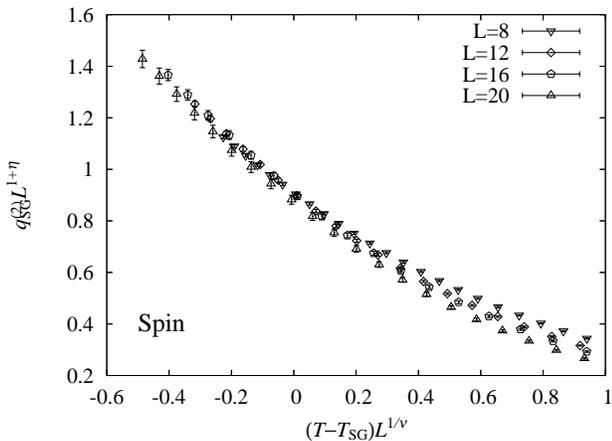}
\caption{
Finite-size-scaling plot of the squared spin-glass order parameter,
assuming a simultaneous chiral-glass and spin-glass transition
with a common correlation length exponent, {\it i.e.\/}, 
$T_{\rm SG}/J=0.19$ and $\nu_{\rm SG}=1.2$. 
The best scaling is obtained with $\eta_{\rm SG}=-0.1$. 
}
\label{fig:fss-q-d00-2}
\end{figure}

After establishing the occurrence of a finite-temperature CG transition, 
we next wish to re-examine  via the finite-size scaling analysis the issue
whether the standard SG order occurs at the same
temperature with the CG order or not. 
In Fig.~\ref{fig:fss-q-d00-2}, we show a finite-size scaling plot of the 
SG order parameter $q_{\rm SG}^{(2)}$, 
assuming a simultaneous CG and SG transition
with a common correlation length exponent,
{\it i.e.\/}, we set
$T_{\rm SG}/J=0.19$ and $\nu_{\rm SG}=1.2$. 
Although the data turn out to scale well 
for smaller sizes, a significant deviation from the scaling
is seen for larger sizes and at lower temperatures. 
The quality of the scaling is not improved if one tries to adjust 
$\nu$ to somewhat larger values. A similar poor scaling behavior is also observed 
even when one instead chooses $|1/T-1/T_{\rm CG}|T_{\rm CG}L^{1/\nu}$ 
as the scaling variable, and tries 
to adjust the scaling parameters around $\nu_{\rm SG}=1.4$.
The data for smaller sizes
turn out to scale best with choosing $\eta_{\rm SG}=-0.1$.
These parameter values $\nu_{\rm SG}=1.2$ and $\eta_{\rm SG}=-0.1$ are
close to the values reported by Matsubara {\it et al\/} in 
Ref.~\onlinecite{Matsubara3}.
Hence, for the SG order parameter, 
we have observed a pseudo-critical behavior for smaller sizes, as well as a 
systematic deviation from the scaling for larger sizes.
If the observed
deviations were due to the correction-to-scaling,   the
scaling should be better for larger sizes, which is opposite to our present
observation. Therefore, we do not consider
the apparent scaling obtained for smaller sizes 
with $T_{\rm SG}/J=0.19$ 
to be acceptable as a true
asymptotic scaling. 
\begin{figure}
\includegraphics[width=\figwidth]{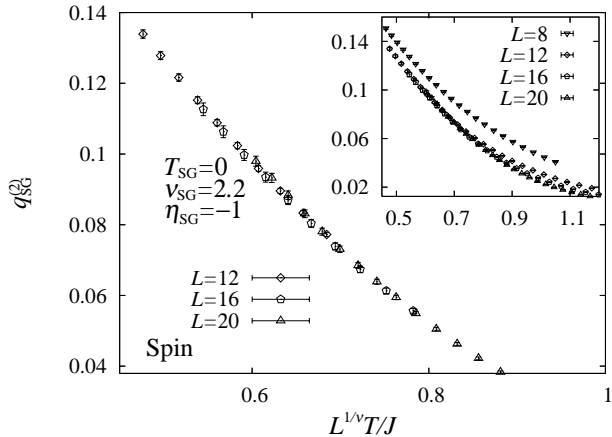}
\caption{
Finite-size-scaling plot of the squared spin-glass order parameter with
assuming a zero-temperature spin-glass transition $T_{\rm SG}=0$. 
In the main panel,
only the data for larger sizes and at lower temperatures, {\it i.e.\/},
those at $T/J\leq 0.19$ and with $L\geq 12$, are plotted. 
In the inset, the same scaling plot using all the data is shown. 
}
\label{fig:fss-q-d00}
\end{figure}

In Fig.~\ref{fig:fss-q-d00}, we show a finite-size-scaling plot of the
SG order parameter
$q_{\rm SG}^{(2)}$ using the same data as in Fig.~\ref{fig:fss-q-d00-2}, 
but  now assuming a zero-temperature SG transition, 
{\it i.e.\/},
$T_{\rm SG}=0$ and $\eta_{\rm SG}=-1$. The
value  $\eta=-1$ is generically expected for a
zero-temperature transition
with the non-degenerate ground state.    
As shown in Fig.~\ref{fig:fss-q-d00}, the best data collapse is obtained
by choosing $\nu_{\rm SG}=2.2$.
If one uses in the scaling plot the data at low temperatures,
lower than the CG transition temperature $T/J\leq 0.19$, 
and the data for larger sizes $L\geq 12$, 
the scaling turns out to work well: See the main panel. By contrast,
If one includes in the scaling plot the data for the smallest size $L=8$ and
at high temperatures $T/J\geq 0.19$, a significant deviation
from the scaling is observed for these data: See the inset. 
In sharp contrast to the scaling plot of Fig.~\ref{fig:fss-q-d00-2} with 
$T_{{\rm SG}}=T_{{\rm CG}}$, 
we have observed here 
a better scaling for larger sizes, and a systematic deviation from the
scaling for smaller lattices. 
In that sense,
the present finite-size scaling 
analysis is fully consistent with the occurrence of 
a $T=0$ 
SG transition, as has long been believed in the community.\cite{Banavar,
McMillan,Olive86,Matsubara0,Yoshino1,Kawamura92,Kawamura98,HukuKawa00a}
Furthermore, the exponent associated with the possible $T=0$ SG
transition happens to be rather close to the previous estimates
based on the numerical domain-wall method.\cite{Banavar,McMillan,Kawamura92} 

Of course, as discussed above, the CG
transition occurring
at $T/J\simeq 0.19$ would necessarily affect the nature of the SG
ordering, even if the Heisenberg
spin itself does not order at $T=T_{\rm CG}$. Thus,
even if the SG transition occurs only at $T=0$, an intrinsic 
SG critical phenomenon associated with this $T=0$ transition
should set in at low temperatures below $T_{\rm CG}$, whereas
the data at and above $T_{\rm CG}$ would be ``contaminated'' by
the CG transition which might well 
change the ordering behavior of the Heisenberg
spin via the associated phase-space narrowing.

Hence, although our present data are fully  consistent with the occurrence of
the $T=0$ SG transition, in order to see such a behavior clearly, 
one has to
choose the scaling region carefully. Inclusion of the data of smaller
sizes and at higher temperatures in the analysis 
would easily deteriorate
the quality of the scaling plot, leading to the opposite conclusion.
We believe that this is indeed the situation of the recent study of 
Ref.~\onlinecite{Matsubara3}, 
in which a simultaneous spin and chiral transition at a finite
temperature $T_{\rm SG}=T_{\rm CG}$ was concluded.

\subsection{Correlation length}
\label{subsec:clength}
\begin{figure}
\includegraphics[width=\figwidth]{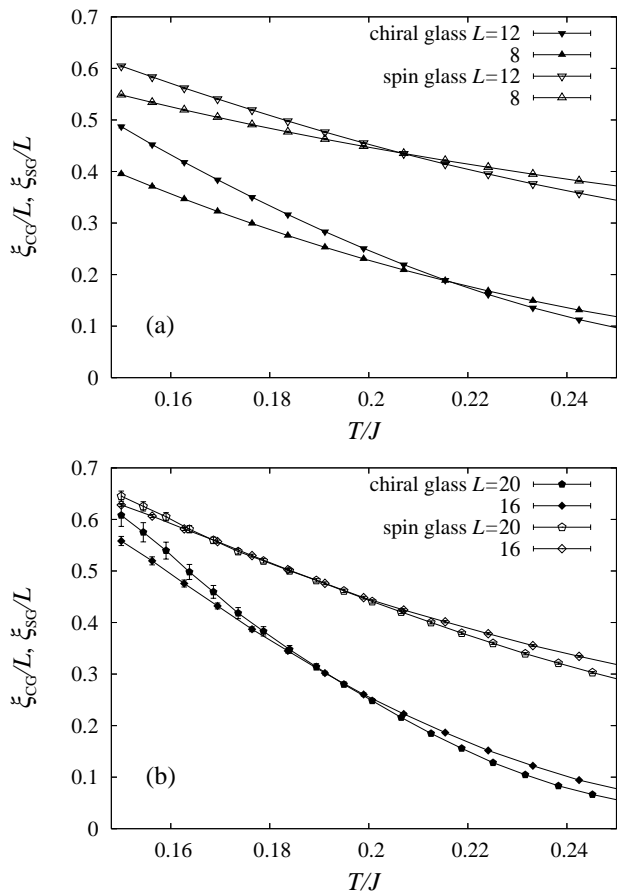}
\caption{
The correlation length divided by the linear size $L$ 
plotted against the   temperature for the chiral-glass  
$\xi_{\rm CG}/L$ (filled symbols)
and for the spin-glass  $\xi_{\rm SG}/L$ (open symbols). 
The data for smaller sizes ($L=8$ and $12$) are shown in the upper figure,
and those for larger sizes ($L=16$  and $20$) are shown in the lower figure.
}
\label{fig:corr-len}
\end{figure}
The temperature dependence of the normalized SG and CG
correlation lengths, 
$\xi_{\rm SG}/L$ and $\xi_{\rm CG}/L$,
for various sizes are shown in Fig.~\ref{fig:corr-len}.  
In contrast to the Binder parameter shown in Fig.~\ref{fig:bin-d00}, 
the normalized correlation lengths of $L=8$ and 12 shown in 
in Fig.~\ref{fig:corr-len}(a)
exhibit a clear crossing
at a temperature around
$T/J\simeq 0.2$ for both cases of the SG and the CG. 
The observed behavior
is consistent with the behavior recently reported  
by Lee and Young\cite{LeeYoung} for the $3D$
Heisenberg SG model with the Gaussian coupling
for the sizes up to $L=12$. 

We now extend the system size up to $L=20$, and the result is presented in
Fig.~\ref{fig:corr-len}(b). While 
the crossing temperature for larger sizes $L=16$ and 20
shifts toward lower temperature for
both cases of the SG and the CG, the CG
correlation length still has  a clear
crossing around $T/J\simeq 0.19$, very close to the estimate of
$T_{\rm CG}$ in the previous subsections, {\it with a finite
crossing angle\/}. 
On the other hand, for the SG correlation length, the crossing 
becomes weaker and almost fades away. Namely,
the curves of $L=16$ and 20
{\it merge nearly tangentially 
with a vanishing crossing angle\/}. The $L=16$ 
and 20 curves of $\xi_{{\rm SG}}/L$ 
stay on top of each other in an entire temperature region studied 
below $T/J\simeq 0.19$, as if they were in the critical 
KT-like phase. Hence, for the SG,
with increasing $L$, not simply 
the crossing temperature shifts toward lower temperature, but 
the crossing-angle becomes smaller and almost vanishes. 
This is in contrast to the
behavior of the CG correlation length where the crossing-angle remains finite
with increasing $L$.
It thus seems possible that the further increase of $L$ eventually leads to
the disappearance of the  crossing for $\xi_{{\rm SG}}/L$, at least
in the temperature range studied here.

This would be consistent with the size-crossover
expected from the spin-chirality coupling/decoupling picture, 
and with
our observation in Sec.~\ref{sec:q2} that the decoupling length scale is
about $L=20$. Unfortunately, at present,
we cannot go to lattices larger than $L=20$ 
due to the
limitation of our computation capability. We certainly expect, however, that
the crossing of $\xi_{{\rm SG}}/L$ eventually disappears, 
or at least shifts to a temperature
considerably lower than the CG transition temperature 
$T_{{\rm CG}}/J\simeq 0.19$, if we could study lattices considerably
larger than $L=20$. For now,
we only mention that, although the recent
data of the normalized correlation length for smaller lattices of $L\leq 12$ 
might look rather conclusive at first sight,\cite{LeeYoung} 
in view of our present data for larger sizes presented in 
Fig.~\ref{fig:corr-len}(b),
it is still difficult to draw a definite conclusion about
the ordering nature of the model only through the correlation-length 
measurements.

\subsection{Overlap distribution}
\label{sec:pofq}

\begin{figure}
\includegraphics[width=\figwidth]{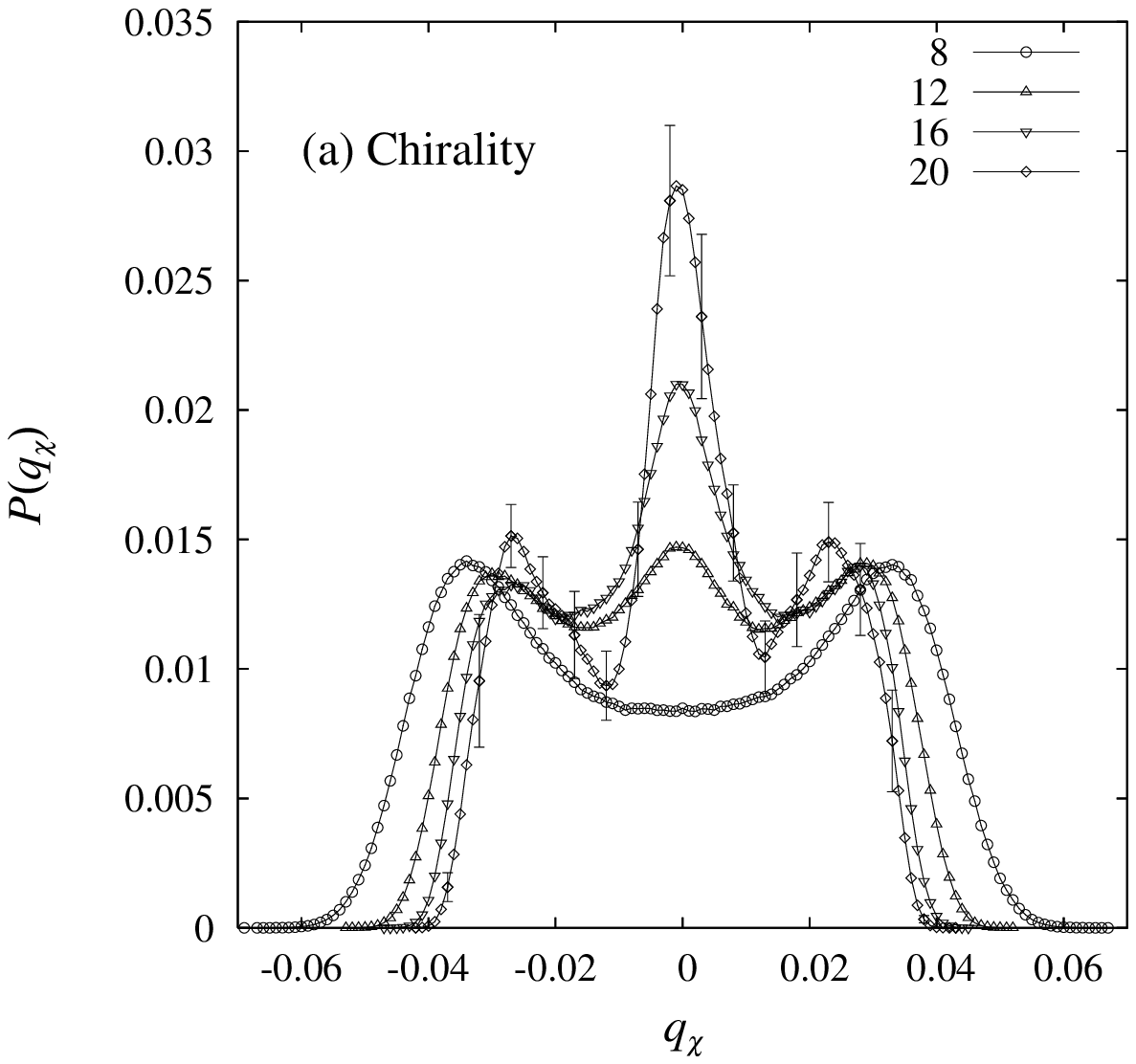}
\includegraphics[width=\figwidth]{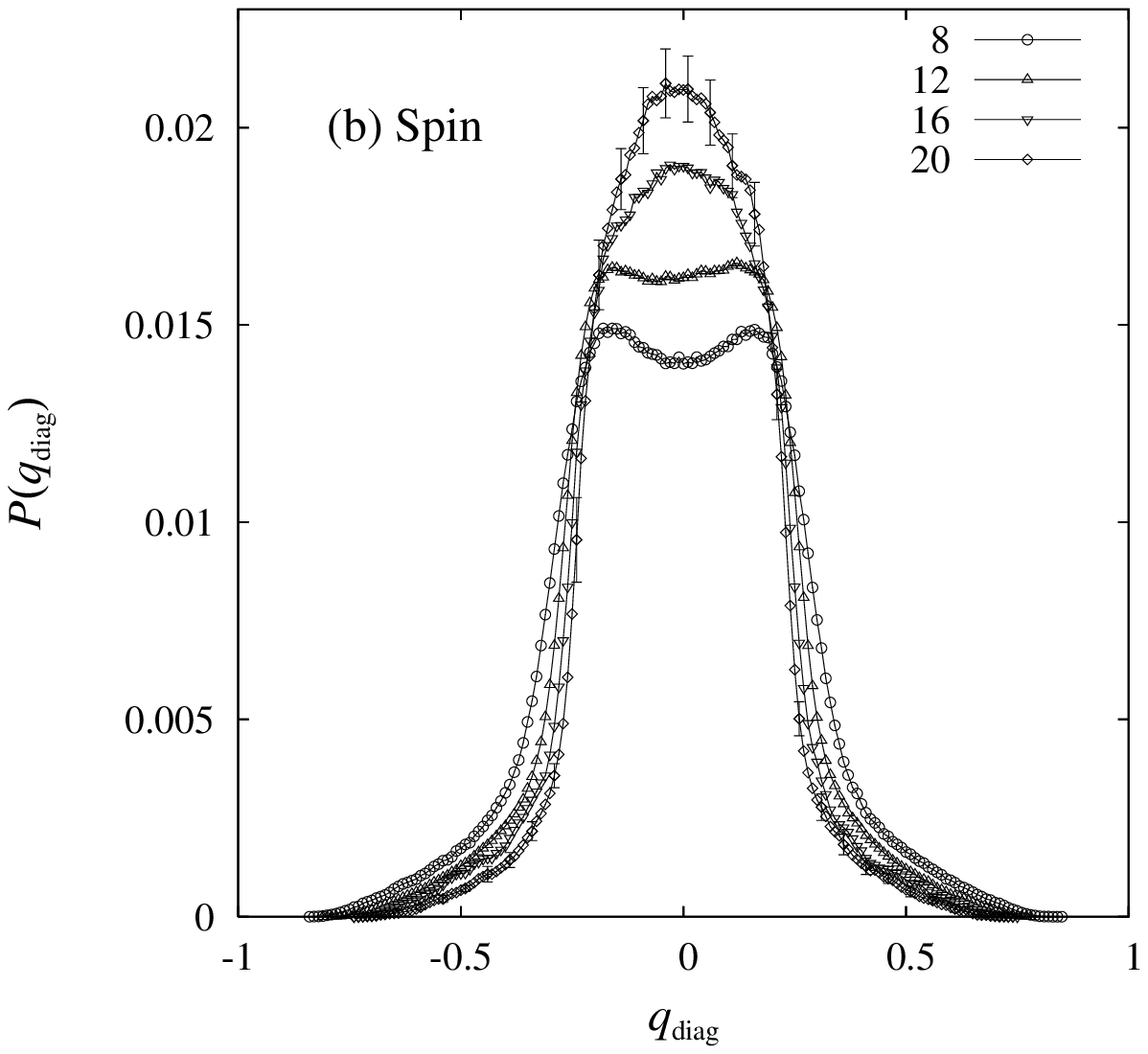}
\caption{
The size dependence of the chiral-overlap distribution function;
upper figure (a), and of the
diagonal-spin-overlap distribution function; lower figure (b), 
at the lowest temperature of the present simulation,
$T/J=0.15$. 
}
\label{fig:pofq-d00}
\end{figure}

In Fig.~\ref{fig:pofq-d00}, we show the chiral-overlap 
distribution function; upper figure (a), 
and the diagonal-spin-overlap distribution function; lower figure (b),
at a temperature $T/J=0.15$ 
below the CG transition temperature.
One sees from Fig.~\ref{fig:pofq-d00}(a) 
of the chiral-overlap distribution function 
$P(q_\chi)$ that,
with increasing $L$, the side peaks
corresponding to the CG EA order parameter $q_{\rm CG}^{\rm EA}$ 
grow and sharpen, which indicates the
occurrence of the CG long-range order. In addition, a central peak
at $q_{\chi}=0$ shows up for $L\geq 12$,  which also grows and sharpens
with increasing $L$. The existence of this central peak coexisting with the
side peaks suggests the occurrence of a one-step-like RSB
in the CG ordered state. This feature is also consistent with
the existence of a negative dip in the CG Binder parameter $g_{\rm CG}$ and 
with the crossing of $g_{\rm CG}$ occurring on the negative side, 
as was discussed in Sec.~\ref{subsec:BP}.
The behavior of $P(q_\chi)$ observed here
is similar to the previous reports for
the 3D Heisenberg SG with the Gaussian coupling\cite{HukuKawa00a}
and the related Heisenberg SG
models.\cite{KawamuraImagawa01,ImagawaKawamura02,ImagawaKawamura03}  
By contrast, such a one-step-like feature of the overlap distribution
has never seen in the Ising SGs both with the short-range 
\cite{Marinari98b} and infinite-range \cite{BhattYoung89} interactions, 
nor in the Heisenberg SG with the infinite-range interaction
\cite{ImagawaKawamura03}. 

Fig.~\ref{fig:pofq-d00}(b) represents the size dependence of
the diagonal-spin-overlap distribution function $P(q_{\rm diag})$
defined by Eq.~(\ref{eqn:dist-q}). For larger $L\geq 16$,
the distribution function $P(q_{\rm diag})$ has only a single peak
at $q_{\rm diag}=0$, which grows with
increasing $L$, without any other divergent peak.
This is in sharp contrast to
the triple-peak structure  observed 
in the chiral-overlap distribution function $P(q_\chi)$
of Fig.~\ref{fig:pofq-d00}(a), peaked at
$q_\chi=0$ and $\pm q_{{\rm CG}}^{\rm EA}$. 
It is also in contrast to the
double-peak structure
observed in $P(q_{\rm diag})$ of the mean-field Heisenberg SK model, peaked at
$q=\pm \frac{1}{3}q_{\rm EA}$.\cite{ImagawaKawamura03}
As discussed in Sec.~\ref{sec:pq}, 
since the diverging peaks at 
$q_{\rm diag}=\pm \frac{1}{3}q_{\rm EA}$ should arise
in $P(q_{\rm diag})$ in the possible SG ordered state with a non-zero
EA SG order parameter,\cite{ImagawaKawamura03}
the absence of any divergent peak 
at non-zero $q_{\rm diag}$ for larger $L$
strongly suggests that the system is in the SG {\it disordered\/} state even
at this low temperature $T/J=0.15$.
Interestingly, a closer inspection of
Fig.~\ref{fig:pofq-d00}(b) reveals that 
a weak double-peak structure can be seen for smaller sizes corresponding to
the spin-chirality coupling regime, 
$L=8$ and $12$. However, such a double-peak
structure in $P(q_{\rm diag})$
tends to disappear for larger sizes corresponding to the
spin-chirality decoupling regime, $L\geq 16$. Again,
this could be interpreted as the size-crossover from the small-size
pseudo SG order to the large-size SG disorder, as is 
naturally expected from the
spin-chirality coupling/decoupling picture.

\subsection{Equilibrium auto-correlation functions}

\begin{figure}
\includegraphics[width=\figwidth]{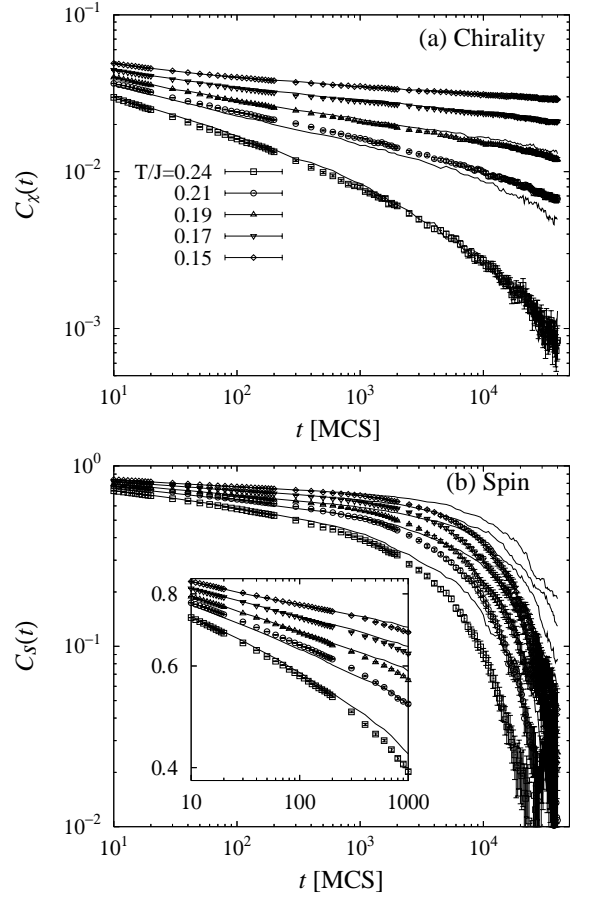}
\caption{
The Monte Carlo time dependence of the chiral autocorrelation function
$C_\chi(t)$;
upper figure (a), and of the spin  autocorrelation
function $C_s(t)$; 
lower figure (b), at various temperatures both below and above
$T_{\rm CG}/J\simeq 0.19$. The system size is $L=16$ (given by symbols) and
$L=20$ (given by thin lines).  
The inset is an enlarged view of $C_s(t)$ in the short-time
region where the finite-size effect is negligible. 
}
\label{fig:cg-c-d00}
\end{figure}

Next, we discuss the ordering behavior of the model
by studying its equilibrium dynamics. In Fig.~\ref{fig:cg-c-d00}, we show
the MC time dependence 
of the chiral and spin
autocorrelation functions for our two largest sizes $L=16$ and $20$. 
Here, the time is measured in units of the 
standard heat-bath MC steps without the temperature-exchange procedure.

In the chiral autocorrelation function $C_{\chi}(t)$ shown 
in Fig.~\ref{fig:cg-c-d00}(a),
no appreciable difference is observed between the data of $L=16$ and 20
in the time window of $t\leq 10^4$, beyond which a weak size effect 
is appreciable. 
The spin autocorrelation function $C_{s}(t)$, by contrast, 
is more susceptible to the 
finite-size effect, as can be seen from Fig.~\ref{fig:cg-c-d00}(b). 
Even in this case, however, 
the data in the time window $t\leq 10^3$ shows a
negligible size effect as shown in the inset.  

\begin{figure}
\includegraphics[width=\figwidth]{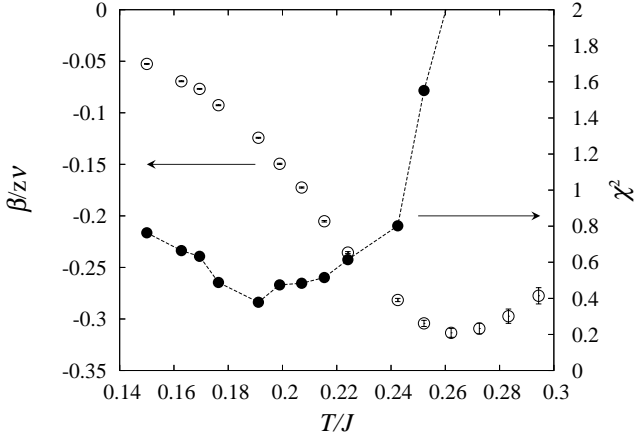}
\caption{
Residuals per degrees of freedom associated with the $\chi^2$-fitting of the
chiral autocorrelation function (marked by filled circle), 
and an estimated effective exponent
 $\beta/z\nu$ (marked by open circle) plotted against the temperature.  
}
\label{fig:fitting-c2}
\end{figure}

As can be seen from Fig.~\ref{fig:cg-c-d00}(a), 
$C_{\chi}(t)$ shows a downward trend above $T/J=0.19$, an 
upward trend below $T/J=0.19$, and a near linear behavior at $T/J=0.19$. 
In order to quantify this, we fit the data at each temperature
to the form (\ref{eqn:qoft-tc}) and plot 
the $\chi^2$-deviation of the fit in Fig.~\ref{fig:fitting-c2} 
as a function of the temperature around the expected CG
transition temperature.  
The plot has a minimum around $T/J=0.19(1)$, at which the data
are optimally fitted to a power law. 
This estimate of $T_{\rm CG}$ based on the chiral autocorrelation function
agrees well with those obtained from the CG Binder parameter
and the CG order parameter. 

\begin{figure}
\includegraphics[width=\figwidth]{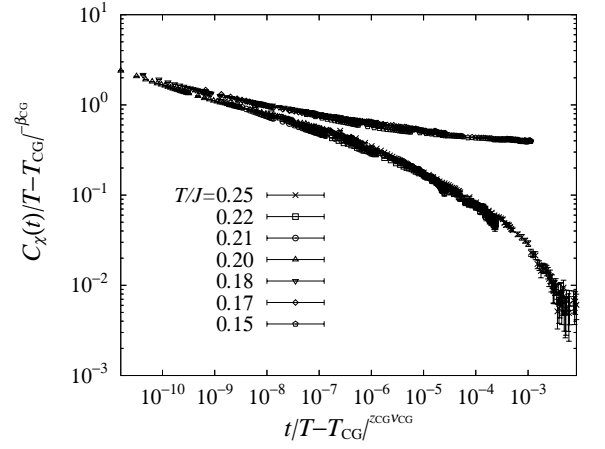}
\caption{
Dynamical scaling plot of the chiral autocorrelation function. The
best scaling is obtained with choosing $T_{\rm CG}=0.19$,
$\beta_{\rm CG}=0.8$ and $z_{\rm CG}\nu_{\rm CG}=5.4$. 
The upper branch represents the scaling of the data below $T_{\rm CG}$, 
while the lower branch represents that above $T_{\rm CG}$.  
}
\label{fig:dynamical-scaling}
\end{figure}

We also test a dynamical scaling analysis of the chiral autocorrelation
function $C_{\chi}(t)$. As can be seen from
Fig.~\ref{fig:dynamical-scaling}, the dynamical scaling  works
well both above and below $T_{\rm CG}$, with the scaling parameters
$T_{\rm CG}/J=0.195$,  
$\beta_{\rm CG}=0.8$ and $z_{\rm CG}\nu_{\rm CG}=5.0$. 
The present estimate of $\beta_{\rm CG}$ is slightly smaller than, but
is not far from the previous estimate of Ref.~\onlinecite{Kawamura98} for the
3D Heisenberg SG with the Gaussian coupling $\beta_{\rm CG}\simeq 1.1$.

As can be seen from
Fig.~\ref{fig:cg-c-d00}(b), by contrast, 
the spin autocorrelation $C_{s}(t)$
shows a downward trend at longer times at any temperature studied, 
suggesting an
exponential-like decay characteristic of the disordered phase.  
A closer inspection of the data of $C_{s}(t)$
reveals that the data below $T_{{\rm CG}}$
exhibit a weak hump-like structure {\it at short times} $t\simeq 10^2$, though
this hump eventually gives way to the down-bending trend characteristic
of the disordered phase at longer times $t\agt 10^3$: See the inset of 
Fig.~\ref{fig:cg-c-d00}(b). This  hump-like structure 
observed in $C_{s}(t)$
at short times might be a manifestation of the trivial spin-chirality coupling 
expected 
at short time scales, and {\it is not likely to be an 
indication of the SG long-range 
order\/}, since the downward trend is recovered at longer time scales:
See Sec.~\ref{sec:discussions}  for further details.
Hence, from our dynamical data, we conclude again that the CG
transition occurs at $T_{\rm CG}/J=0.19(1)$, without accompanying the
simultaneous SG order. 

\subsection{Equilibrium correlation time}
\label{sec:tau}

\begin{figure}
\includegraphics[width=\figwidth]{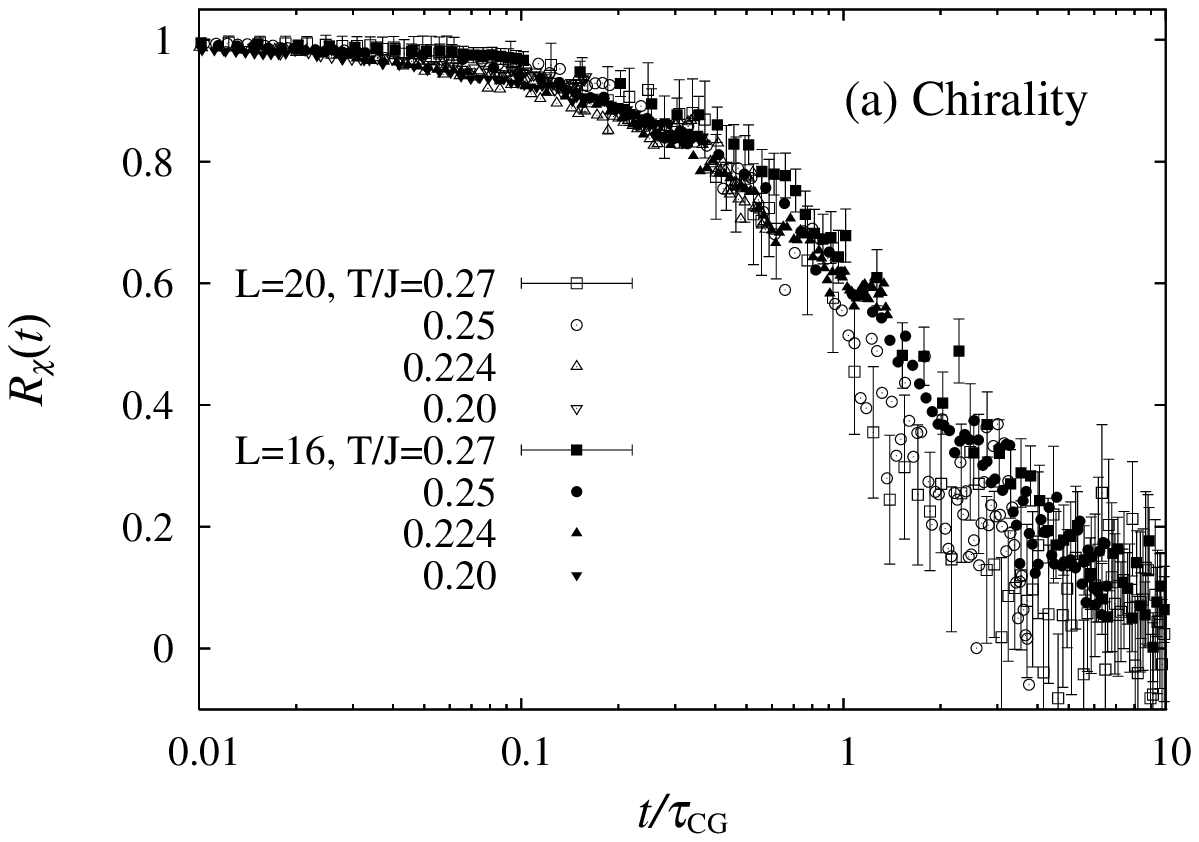}
\includegraphics[width=\figwidth]{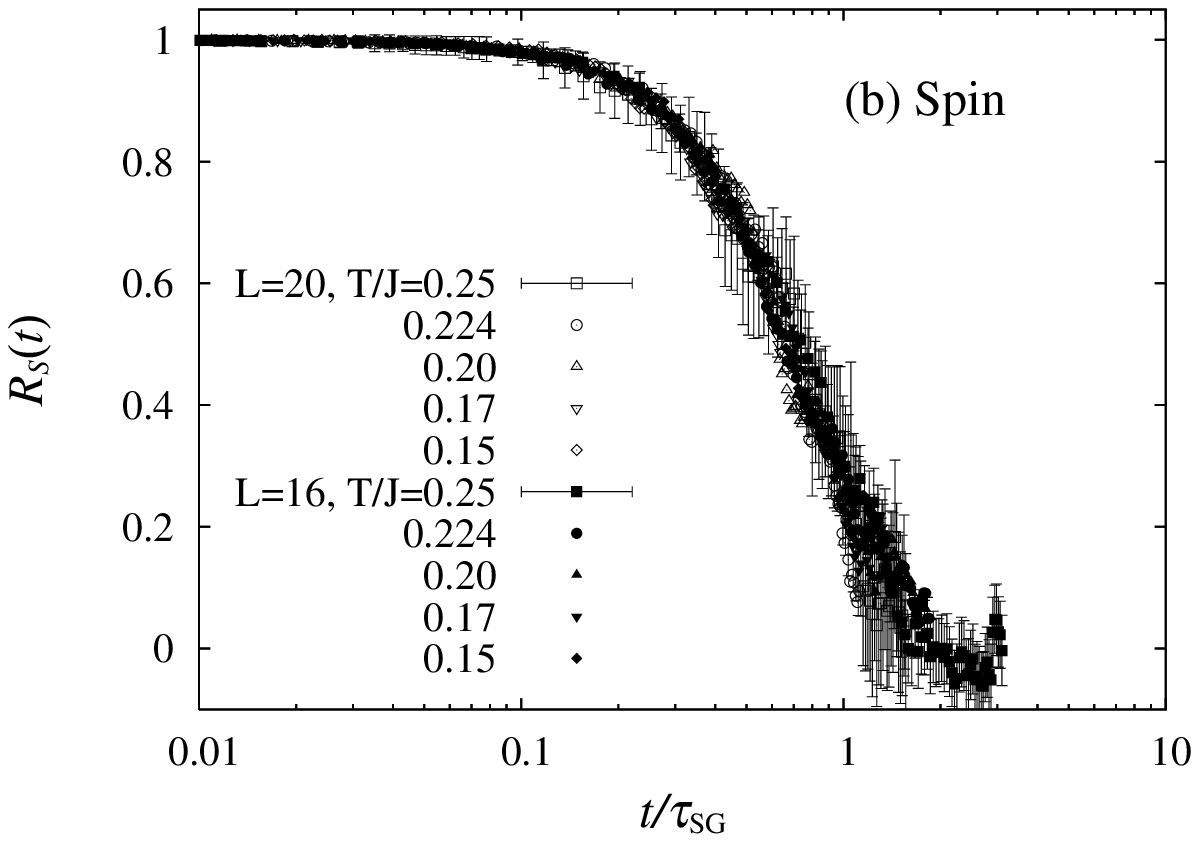}
\caption{
Scaling plot of the 
chiral ratio function; upper figure (a), and of the spin ratio function;
lower figure (b),  at various temperatures for the sizes $L=16$ and $20$. 
}
\label{fig:cg-c-rs-d00}
\end{figure}

Generally, the temporal decay of the autocorrelation is characterized by 
the temperature-dependent characteristic time scale, 
the correlation time $\tau(T)$, which represents a dynamical crossover
from the short-time critical behavior to the long-time relaxation. 
The correlation time $\tau(T)$ diverges as
$\tau(T)\simeq |T-T_{\rm c}|^{-z\nu}$ when the temperature approaches
$T_{\rm c}$ from above.  
One promising method of estimating $\tau$ from the autocorrelation
function has been proposed by Bhatt and Young\cite{BhattYoung92}, who
employed a scaling analysis of the dynamical ratio function. For
the spin autocorrelation, this reads as
\begin{equation}
R_s(t)   =  \frac{C_s(t)}{
\sqrt{
\frac{1}{N}\left[
\left\langle
\left(\sum_i \vec S_i(t_0)\cdot \vec S_i(t+t_0)\right)^2
\right\rangle\right]
}
}. 
\end{equation}
The corresponding chiral ratio function is defined by 
\begin{equation}
  R_\chi(t) = \frac{C_\chi(t)}{
\sqrt{
\left[
\left\langle \left(\frac{1}{3N}\sum_{i\mu}\chi_{i\mu}(t_0)\chi_{i\mu}(t+t_0)\right)^2
\right\rangle\right]
}
}.
\label{eqn:ratio}
\end{equation}
Because the ratio function is dimensionless, the prefactor
$t^{-\beta/z\nu}$ in Eq.~(\ref{eqn:dss}) is canceled out.
The dynamical scaling form of $R(t)$ is then given as a single-variable
function of $t/\tau$,  
\begin{equation}
  R(t)=\overline{R}(t/\tau),
\end{equation}
where $\overline{R}$ is a scaling function. 
If one appropriately chooses the scaling parameter $\tau$ which depends on
the temperature and the system size, the ratio functions should be 
scaled on to a single curve. 
Using this method, Bhatt and
Young\cite{BhattYoung92} successfully estimated the correlation time of 
the short-range Ising EA model and the mean-field Ising SK model. 
Subsequently, this method has been extended to non-equilibrium
relaxation,  where the ratio function depends not only 
on the measurement time $t$ but also on the waiting time $t_w$.
The off-equilibrium 
method was applied recently by Matsumoto, Hukushima and Takayama
to the $3D \pm J$  Heisenberg SG.\cite{Matsumoto}

Here we use this method to estimate the correlation
times both for the spin and for the chirality. 
In comparison with the previous off-equilibrium study,\cite{Matsumoto}  
the present equilibrium study has an advantage that one
needs not extrapolate to an equilibrium limit, {\it i.e.\/}, needs not take the
$t_w\rightarrow \infty$ limit.
In Fig.~\ref{fig:cg-c-rs-d00}, we show the scaling plot of the chiral and 
spin ratio functions.  
We note that both the spin and chiral scaling functions are 
described roughly by an exponential form. 
\begin{figure}
\includegraphics[width=\figwidth]{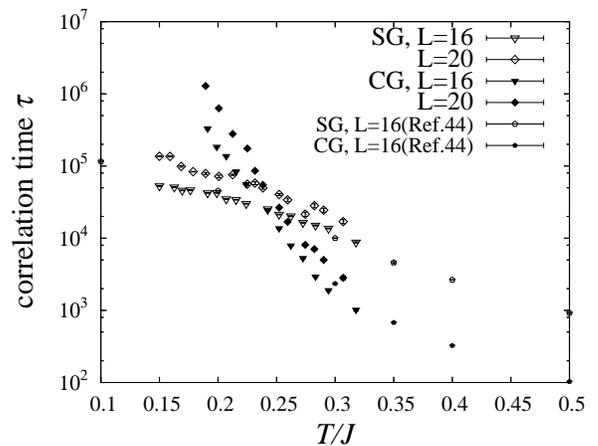}
\caption{
The temperature dependence of the the chiral and spin correlation times 
for the sizes $L=16$ and  $20$. The corresponding data obtained from the
off-equilibrium simulation of Ref.~\onlinecite{Matsumoto} are also included.}
\label{fig:tau-d00}
\end{figure}

\begin{figure}
\includegraphics[width=\figwidth]{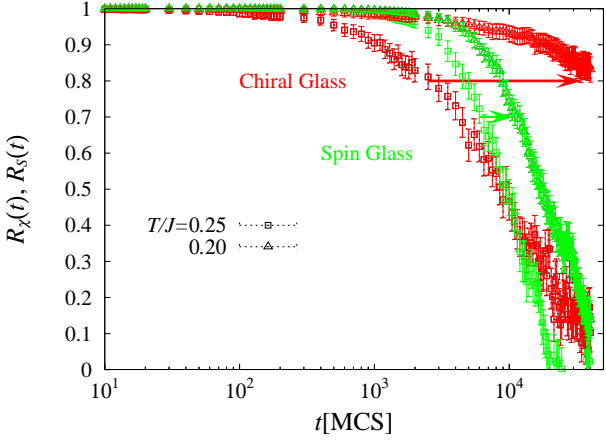}
\caption{
Temporal decay of the chiral and spin ratio functions at 
temperatures $T/J=0.25$ and $0.20$. The system size is $L=16$. 
With decreasing the temperature from $T/J=0.25$ to $0.20$, 
the chiral
relaxation slows down much more slowly than the spin relaxation, as
illustrated by the arrows. }
\label{fig:cq-c-r-d00}
\end{figure}

In order to compare the spin and chiral correlation times,
denoted by $\tau_{\rm SG}$ and $\tau_{\rm CG}$, respectively, 
we plot them in Fig.~\ref{fig:tau-d00} as a function of the temperature.
In the figure, 
we combine the data with those obtained
in a wider temperature range by off-equilibrium simulation of 
Ref.~\onlinecite{Matsumoto}. 
As can be seen from Fig.~\ref{fig:tau-d00}, 
the chiral correlation time is shorter than the spin
correlation time at higher temperatures, similarly to the behavior of the correlation length 
discussed in Sec.~\ref{subsec:clength}. 
As the temperature is decreased, the chiral
correlation time $\tau_{\rm CG}$ grows faster than the spin correlation time 
$\tau_{\rm SG}$, 
and eventually exceeds
$\tau_{\rm SG}$ at a certain characteristic temperature $T=T_\times(L)$. 
The size dependence of this crossover temperature
$T=T_\times(L)$ is apparently weak: We get
$T=T_\times\simeq 0.24$ both for $L=16$ and 20. 
It strongly suggests that, even in the thermodynamic limit,
the chiral correlation time exceeds the spin correlation time at a 
crossover temperature $T_\times (\infty)$, which is located somewhat 
above the CG transition temperature $T_{{\rm CG}}/J\simeq 0.19$. 
It means that, with decreasing the temperature, 
the relevant degree of freedom dominating the
long-time ordering behavior changes from the spin to the chirality at 
$T=T_\times$. In order to further illustrate this changeover, 
we show  
in Fig.~\ref{fig:cq-c-r-d00} the time-dependence of the spin and 
chiral ratio functions at two representative temperatures
$T/J=0.25$ and 0.20, each above
and below $T_\times$. As can be clearly seen from the figure,
with decreasing the temperature across $T_\times$, the temporal decay of
the chiral ratio function becomes much slower than that of the spin
ratio function: Compare the two arrows in the figure.

The time scale associated with such a crossover, $t_\times$, is roughly
estimated to be $10^5\sim 10^6$ MCS. For more
precise estimate of $t_\times$, more quantitative analysis of the
size dependence of the crossover time scale would be necessary. 
Naturally, this crossover time
$t_{\times}$ gives a measure of the time scale above which 
the spin-chirality decoupling can be
observed in dynamics. Thus, the spin-chirality decoupling 
in the dynamics would be eminent
only at temperatures lower than $T_\times/J\simeq 0.24$ 
and at times longer than
$t_\times \simeq 10^5-10^6$ MCS. 
This crossover time scale is rather
long, yet, is finite. It is important to realize that, 
in order to resolve the controversy concerning the
presence or the absence of the spin-chirality decoupling in the
Heisenberg SG, one has to probe the {\it equilibrium\/} 
dynamics beyond this crossover time scale $10^5-10^6$ MCS at temperatures
lower than $T_\times /J\simeq 0.24$, about some $20\sim 30$ \% 
above $T_{{\rm CG}}$.\cite{Picco04} 

\begin{figure}
\includegraphics[width=\figwidth]{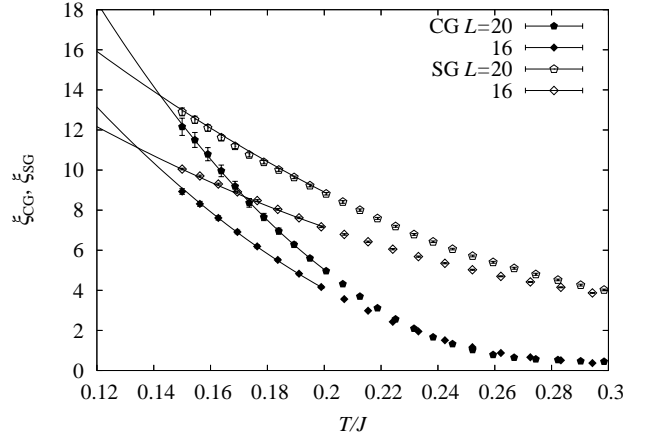}
\caption{The temperature dependence of the chiral and spin 
correlation lengths for finite systems. The system size is $L=16$ and 20.
The data are the same as those shown
in Fig.~\ref{fig:corr-len}, but not divided by
$L$ here. The curves are polynomial fits of the data which are extrapolated
to lower temperatures to deduce the crossing temperature given in the text. }
\label{fig:xi-d00}
\end{figure}

As argued in Sec.~\ref{sec:background}, from the spin-chirality
coupling/decoupling picture,  
a similar phenomenon is expected
also in the spatial correlation of the model
in terms of the length scale.
Namely, one expects that at a certain crossover temperature $T'_\times$,
which is probably close to the dynamical crossover temperature
$T_\times$ discussed above, 
the CG correlation length $\xi_{{\rm CG}}$ exceeds the SG correlation
length $\xi_{{\rm SG}}$. This changeover of the dominant length scale gives a  
crossover length scale $L_{\times}$ above which the spin-chirality decoupling 
is
eminent in spatial correlations. Unfortunately, unlike the case of the
correlation time, the limitation of the available system size prevents us
from directly estimating $L_\times$.
In Fig.~\ref{fig:xi-d00}, we plot the temperature dependence of the
CG and SG correlation lengths for the sizes $L=16$ and 20.
For these sizes, the crossing of $\xi_{{\rm SG}}$ and $\xi_{{\rm CG}}$ occurs 
at a temperature 
lower than the CG transition temperature, in contrast to the case
of the correlation time. Nevertheless, the crossover temperature at which 
$\xi_{{\rm SG}}$ and $\xi_{{\rm CG}}$ of finite $L$ cross, 
tends to increase with increasing $L$.
If we roughly estimate the crossover length scale of finite
systems by extrapolating the data of Fig.~\ref{fig:xi-d00}, we tentatively
get $L_\times\simeq 11$ ($L=16$)
and $L_\times\simeq 14$ ($L=20$). These results are certainly not
inconsistent with our estimate of $L_\times\simeq 20$
based on the behaviors of the SG order parameter, 
the dimensionless correlation length $\xi/L$ 
and other quantities.

\subsection{The $A$ and $G$ parameters}

\begin{figure}
\includegraphics[width=\figwidth]{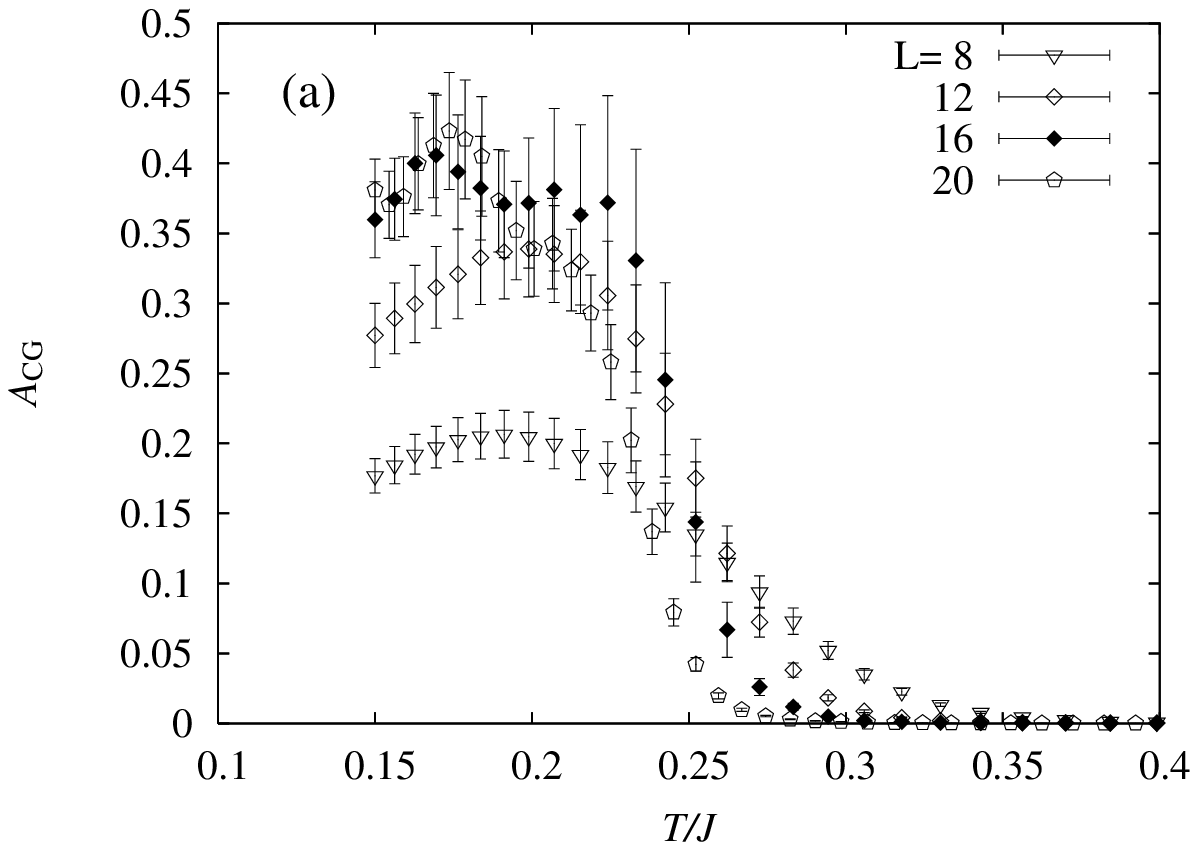}
\includegraphics[width=\figwidth]{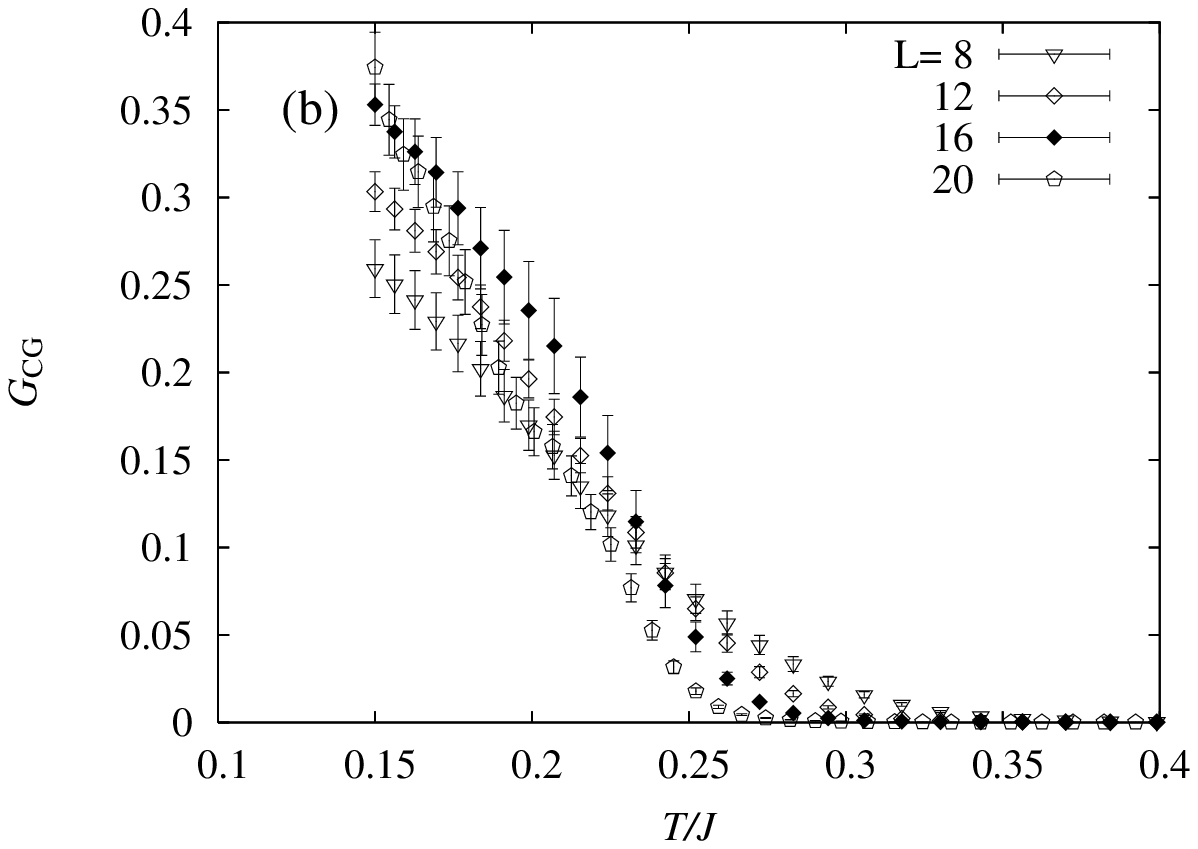}
\caption{
The temperature and size dependence of the $A$ parameter of the chirality; 
upper figure (a), and 
of the $G$ parameter of the chirality; lower figure (b).
}
\label{fig:GA-chiral}
\end{figure}
\begin{figure}
\includegraphics[width=\figwidth]{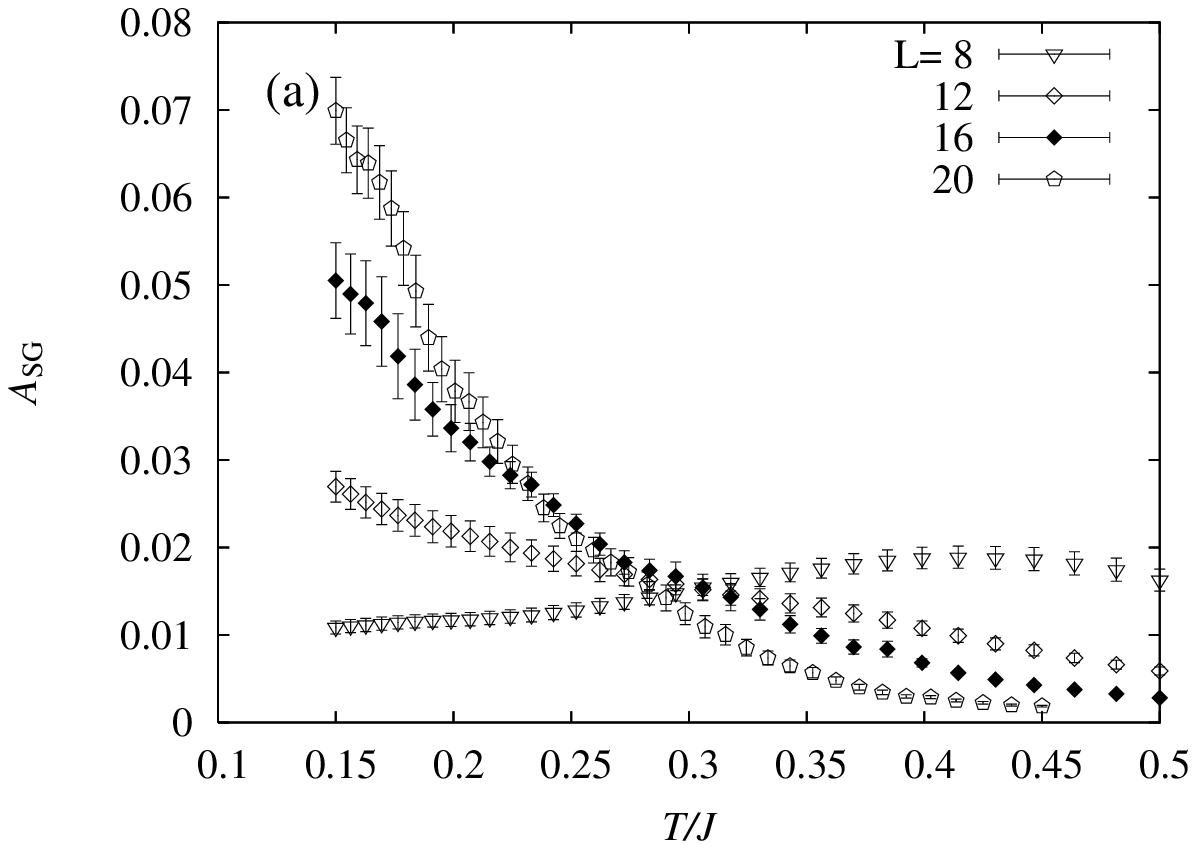}
\includegraphics[width=\figwidth]{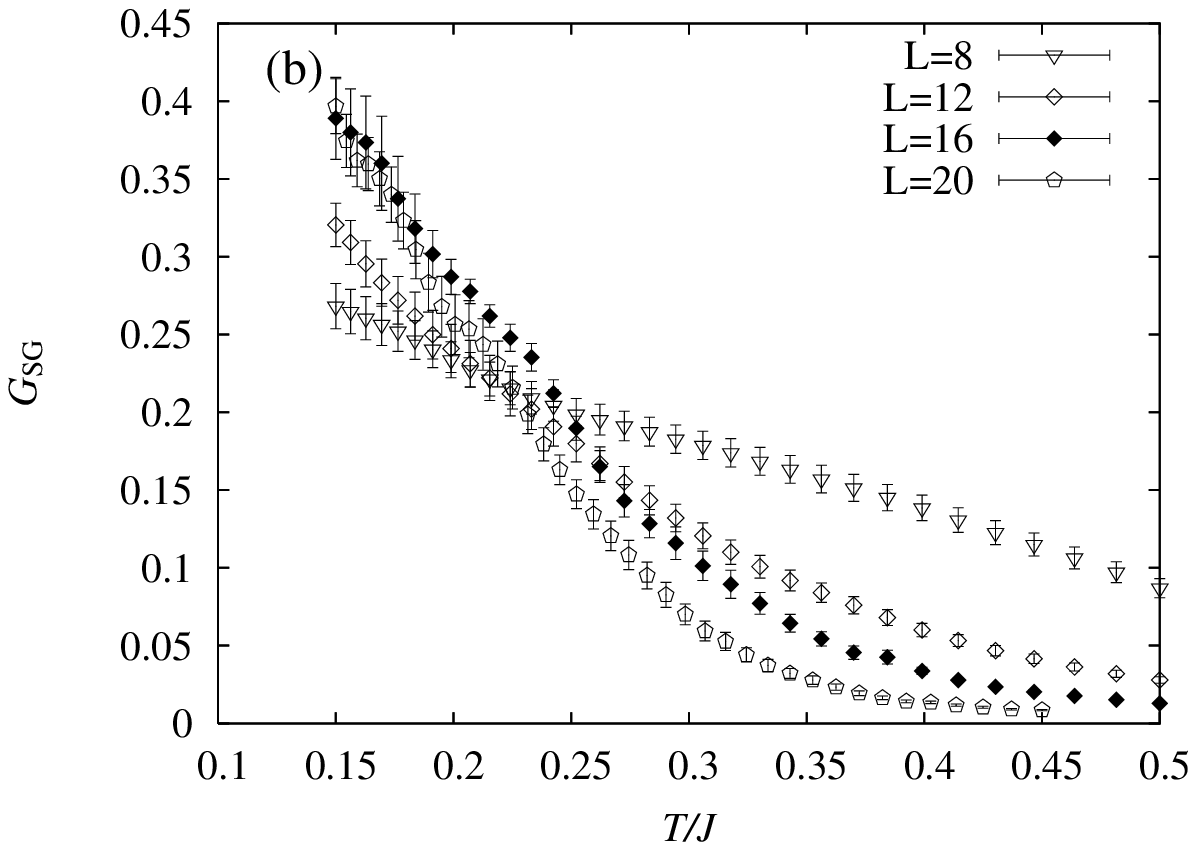}
\caption{
The temperature and 
size dependence of the $A$ parameter of the spin; upper figure 
(a), and 
of the $G$ parameter of the spin; lower figure (b). 
}
\label{fig:GA-spin}
\end{figure}

We have also calculated the $A$ and $G$ parameters defined in Sec.~\ref{sec:pq}
both for the CG and SG orders.
In Fig.~\ref{fig:GA-chiral},
the temperature and size dependence of the
$A$ and $G$ parameters for th CG order, $A_{{\rm CG}}$ and $G_{{\rm CG}}$, 
is shown. 
Although the data are rather noisy due to the large sample-to-sample 
fluctuations,  
the $A$ parameter of different $L$
show a crossing and a peak around the expected CG transition point
$T/J\simeq 0.19$, as can be seen from Fig.~\ref{fig:GA-chiral}(a). 
In particular, with increasing  $L$, $A_{{\rm CG}}$ stays non-zero
below $T_{\rm CG}$,  indicating that the 
CG ordered state is non-self-averaging.
These findings combined with the peculiar shape of $P(q_{\chi})$
shown in Sec.~\ref{sec:pofq} suggest that the CG ordered phase accompanies
an RSB with a non-self-averageness. 
For the corresponding $G$ parameter, the crossing is not so clear, as is 
shown in Fig.~\ref{fig:GA-chiral}(b). 

The temperature and size dependence of the
$A$ and $G$ parameters for th SG order, $A_{{\rm SG}}$ and $G_{{\rm SG}}$, 
is shown in Fig.~\ref{fig:GA-spin}. 
As shown in Fig.~\ref{fig:GA-spin}(a), the $A$ parameter 
exhibits a crossing, although the crossing temperature is located considerably
above $T_c$ for this range of $L$. One might be tempted to interpret 
such a crossing of $A_{{\rm SG}}$ as an unambiguous evidence
of the occurrence of the standard SG transition. 
However, one has to be careful here:
Although the
crossing of $A_{{\rm SG}}$ is certainly a signature of some sort of phase
transition occurring there, 
it does not necessarily mean the occurrence of 
the standard SG transition characterized by a non-zero SG order parameter. 
A non-zero $A_{{\rm SG}}$ persisting in the
$L\rightarrow \infty$ limit simply means that the SG order parameter 
$q_{{\rm SG}}^{(2)}$, or the
SG susceptibility $\chi_{{\rm SG}}=Nq_{{\rm SG}}^{(2)}$, 
is non-self-averaging. 
Below the CG transition temperature, one expects that 
the SG order parameter is still Gaussian-distributed around zero
with a width corresponding to the finite SG susceptibility $\chi_{{\rm SG}}$, 
while the width
exhibits sample-to-sample fluctuations leading to the non-self-averaging 
$\chi_{{\rm SG}}$. 
The latter is a natural consequence of the phase-space
narrowing which should inevitably accompany the CG transition with the
one-step-like RSB. Hence, the crossing of $A_{{\rm SG}}$, and
a finite $A_{{\rm SG}}$ remaining in the $L\rightarrow \infty$ limit
below $T_{{\rm CG}}$, are compatible with the absence of the 
standard SG long-range
order, and is entirely 
consistent with the the CG transition not accompanying 
the standard SG long-range order.

As shown in Fig.~\ref{fig:GA-spin}(b), the $G$ parameter of the spin
exhibits a crossing around $T_{{\rm CG}}$. 
The relation Eq.~(\ref{eqn:GAg}), combined with our  
observation in Fig.~\ref{fig:bin-d00}(b),  
indicates that $G_{{\rm SG}}$ also takes
a non-zero value below $T_{{\rm CG}}$. Thus, the observed
crossing of $G_{{\rm SG}}$
is just as one expects for the CG transition. In other words,
one cannot interpret the crossing of $G_{{\rm SG}}$
as an indicator of the onset of the
standard SG long-range order.

\section{Discussions}
\label{sec:discussions}

In this section, in view of our MC results presented in the previous section,
we wish to examine and discuss
the recent numerical studies on the 3D Heisenberg SG.
Many of these studies suggested, contrary to our present study, 
that the spin and the chirality 
ordered simultaneously at a finite temperature 
with a common correlation length 
exponent $\nu_{{\rm SG}} =\nu_{{\rm CG}}$, {\it i.e.\/}, no spin-chirality
decoupling in the 3D Heisenberg
SG.\cite{Matsubara1,Matsubara2,LeeYoung,BY,Nakamura,Matsubara3}
Below, we wish to make some comments  on these numerical
works from the standpoint of the spin-chirality coupling/decoupling
picture.

\subsection{Stiffness method}

First, we wish to discuss the analyses 
based on the stiffness method.\cite{Matsubara1,Endoh1,Kosterlitz}
In this method, one computes by some numerical means  the change of the
ground-state energy of
finite systems of size $L$ under the appropriate 
change of boundary conditions imposed on the system. This
energy is called a stiffness energy
(or a domain-wall energy), $\Delta E_L$, which gives a measure of an 
energy scale of
low-energy excitations of size $L$.
For large $L$, $\Delta E_L$ is expected to behave as  a power-law, 
$\Delta E_L\approx L^\theta$,
$\theta$ being a stiffness exponent.
If $\theta <0$,
the system remains in the disordered state at any nonzero
temperature, whereas if $\theta >0$
the system possesses a finite long-range order 
at low enough temperatures with $T_c>0$.
Here, we discuss this
method first in conjunction with the detection of the 
standard SG order, leaving the detection
of the CG order later.

The nontrivial part of this stiffness method concerns with the choice
of the boundary conditions employed in computing the stiffness energy. 
There could be various choices, and the behavior of $\Delta E_L$ might 
in principle depend on these choices 
particularly for small sizes accessible in numerical simulations. 
The most standard choice is the combination of 
the periodic and the antiperiodic boundary conditions (P/AP). 
In the case of the Heisenberg SG, the P/AP combination necessarily
accompanies a flipping of the chirality (remember that the 
chirality of the Heisenberg spin is odd
under the spin inversion $\vec S\rightarrow -\vec S$), so that the
P/AP combination should
detect the chiral order for large
enough $L$.\cite{Kawamura92}  
In order to detect the standard
SG order independently of the CG order
by this stiffness method, Ref.~\onlinecite{Kawamura92} introduced
the ``rotation'' boundary conditions (ROT), which imposed a $\pi$ 
rotation on the
boundary spins without flipping the chirality, which was combined with the
standard P boundary conditions in calculating the stiffness 
energy.\cite{Kawamura92}  
Such a  P/ROT combination applied to the 3D Heisenberg SG
yielded a negative $\theta$, {\it i.e.\/}, $\theta\sim -0.51$ 
for the Gaussian coupling, and $\theta\sim -0.49$ for the 
$\pm J$ coupling, which implied the absence of the standard SG order at
nonzero temperature.\cite{Kawamura92} 

By contrast, Matsubara, Endoh and Shirakura
proposed a different choice of boundary conditions in computing $\Delta E_L$,
{\it i.e.\/}, 
to use the free (open) boundary conditions as a reference and impose the
rotational-twist to such ``optimized'' spin configurations 
obtained under the free boundary conditions in which the stress at the boundary
is released.\cite{Matsubara1,Endoh1} 
These authors observed that the stiffness exponent evaluated in this way was
largely positive, close to the spin-wave
exponent $\theta =1$, and argued that the 3D Heisenberg SG 
exhibited 
a finite-temperature SG transition. The method similar in spirit 
to the one used in Refs.~\onlinecite{Matsubara1} and \onlinecite{Endoh1}
was also applied to the {\it XY\/} SG by Kosterlitz and
Akino,\cite{Kosterlitz} leading to the similar conclusion. 
Thus, the result obtained by applying 
the free/twisted-free (F/TF) boundary conditions, 
$\theta >0$ implying $T_{{\rm SG}}>0$,
is in sharp contrast to 
the result obtained by applying the P/ROT boundary conditions, 
$\theta <0$ implying $T_{{\rm SG}}=0$.
Discrepancy between the stiffness exponents evaluated
by the different choices of boundary conditions was also observed
in other SG models, {\it e.g.\/}, in the 2D Ising SG.\cite{Carter}

Although the authors of Refs.~\onlinecite{Matsubara1} and
\onlinecite{Endoh1} argued that their ``optimized'' boundary
conditions were superior to the other choices, 
a theoretical basis of such a claim seems not so obvious.
For example, one might make the following counter-argument
that one should {\it not\/} optimize the boundary conditions
in calculating $\Delta E_L$: 
In the spirit of the domain-wall renormalization-group idea 
by Bray and Moore\cite{Bray}, $\Delta E(L)$ represents an energy scale 
associated with
the interaction between the coarse-grained blocks of size $L$
in an infinite SG sample. Since these
blocks are necessarily subject to the strong
frustration effect caused by the interaction
with the neighboring blocks surrounding them, an optimization of their
energy, independently at each block ignoring the
inevitable frustration effect due to the interaction
with the neighboring blocks, is hardly compatible with the original RG idea.
One may thus argue that, in calculating $\Delta E_L$ in SGs, 
one should  {\it not\/} make the optimization of boundary conditions 
referring to the particular bond realization of each sample.

Concerning the apparent difference of the stiffness exponents
arising from the different choices of boundary conditions, 
there generally exist two possibilities: Either, (i) the observed
difference is a finite-size effect where  
there is a single stiffness exponent for large enough lattices
independent of applied boundary conditions, or, (ii) the
observed difference
is a bulk effect which persists even in large enough lattices.
In the case of the 2D Ising SG, 
Carter, Bray and Moore numerically observed that, although
both the P/AP and F/TF boundary conditions yielded the same
stiffness exponent asymptotically for large $L$, {\it i.e.\/}, 
the possibility (i) above, 
the finite-size effect was much reduced in the P/AP than in the 
F/TF.\cite{Carter}  
For vector SG, there so far exists no convincing evidence 
which of the above (i) and (ii) 
is really the case. In any case, a practical question we face with is 
which set of boundary conditions gives a true asymptotic answer from
smaller sizes accessible in simulations.

One plausible criterion
might be that, among all possible excitations in the system,
the one giving the lowest excitation energy
$\Delta E_L$, or equivalently, the one giving the smallest stiffness exponent 
$\theta$, should be chosen. The reason is simply because
among all possible excitations the one with the
lowest excitation energy should be most efficient in destroying the order
so long as it has non-negligible weight in the thermodynamics,
and would dominate the low-energy dynamics of the model.
Under this criterion, when the different choices of boundary conditions yield
different $\theta$ 
values, the one giving the smallest $\theta$, or the most negative $\theta$, 
should be chosen. In particular, when one set of boundary conditions 
yields a positive $\theta$ while the
other yields a negative $\theta$, the one giving a negative
$\theta$ should be chosen. If so, 
in the case of the 3D Heisenberg SG of our interest, 
the P/ROT combination without any optimization procedure 
should be chosen since it gives the lowest
$\theta$ (negative $\theta$) reported so far.\cite{Kawamura92}  
This suggests that the standard SG order in the 3D Heisenberg SG
occurs only at $T=0$. 
Of course, it is still possible that some other
type of boundary conditions might yield a still smaller $\theta$, 
but it does not 
change the conclusion that the SG order occurs only at $T=0$.

It should also be remembered that 
the types of low-energy excitations generated via a particular choice of 
boundary conditions is only a subset of 
all possible excitations in the system: 
They are basically {\it wall-like\/} excitations, 
not including more complex excitations like, say, 
a ``vortex'' excitation which is possible in 
the Heisenberg SG reflecting the $SO(3)$ nature of its order parameter
space,\cite{comment}  
or a "sponge" excitation which is closely related to
the RSB structure of the ordered state.\cite{Martin,Palassini} 
Unfortunately, 
we have little knowledge 
concerning what is the most relevant low-energy excitation governing the
ordering of the system, and hence, 
have no well-based criterion to choose one
set of boundary conditions  from the others as superior. 
Although we feel that 
our argument above speaks for a
zero-temperature SG transition in the 3D Heisenberg SG, it would be fair to 
say at present that no definitive
conclusion can be drawn  solely based on this stiffness
method.  

We finally wish to 
refer to the stiffness method in detecting the chiral order.
As mentioned, 
since the sign of the chirality is flipped by changing the boundary conditions
from P to AP, the most standard P/AP  
combination could be used in detecting the chiral order, at least for
large enough $L$.  In practice, however, the application of the reflecting
(R) boundary is more efficient in detecting the chiral order, as shown in
Ref.~\onlinecite{Kawamura92}.
The chiral stiffness exponent 
of the 3D Heisenberg SG determined in this way turned out to be positive 
implying a CG transition occurring at a nonzero 
temperature.\cite{Kawamura92} 
Other authors also reported a positive value for the chiral stiffness
exponent both for the 3D {\it XY\/} SG \cite{Kosterlitz} and the 3D 
Heisenberg SG \cite{Endoh1}.

\subsection{Equilibrium dynamics}

Matsubara, Shirakura and Endoh  
reported a further evidence of the simultaneous spin and chiral 
transition  in the 3D Heisenberg SG by investigating the equilibrium 
spin dynamics\cite{Matsubara2}.   
In order to eliminate the effect of global spin rotations inherent to
finite systems, Matsubara {\it et al\/} introduced an artificial
global-rotation correction in the spin dynamics of the model.
They observed that the
modified spin autocorrelations adjusted by the global-rotation correction
exhibited at lower temperatures a tendency to approach a nonzero value 
at longer times, 
which was interpreted as an evidence of  a finite 
SG long-range order.

It should be noticed here that, when one
looks at a quantity which is {\it even\/} under the symmetry transformation 
of the Hamiltonian like the modified spin autocorrelation
function of Ref.~\onlinecite{Matsubara2}, 
one needs to examine its
size dependence carefully. As is well-known,
an even quantity  in finite systems always takes a nonzero value 
even above $T_c$ due to the
finite-size effect, where this nonzero value decreases with the size $L$, 
eventually vanishing  as $L\rightarrow \infty$ above $T_c$.
(Indeed, 
in an extreme occasion of a single spin, the modified
spin autocorrelation function as computed by Matsubara {\it et al\/} does not
decay at all even at an infinite temperature !)
The ordering behavior of the modified spin autocorrelation as
observed by
Matsubara {\it et al\/} might possibly be caused by the 
finite-size effect. In order to refute such  
suspicion, one 
needs to study its size dependence carefully, whereas the analysis of 
Ref.~\onlinecite{Matsubara2} was
limited  to a fixed size $L=16$. 
 
It should be stressed that, 
even within the spin-chirality coupling/decoupling scheme, 
it is still possible that   
the spin autocorrelation function $C_s(t)$ of an infinite
system exhibits  below $T_{{\rm CG}}$ a hump-like weak structure at short times
as illustrated in Fig.\ref{fig:cross-c}, 
which is an echo of the plateau-like structure 
of the {\it chiral\/} autocorrelation function. 
In the temperature range $T_{{\rm SG}}<T<T_{{\rm CG}}$, 
such a hump  of $C_s(t)$ appears only at times shorter than the 
crossover time scale $t_\times$ ($t_\times$ was 
estimated to be $t_\times\approx 10^5-10^6$ above $T_{{\rm CG}}$), while 
$C_s(t)$ eventually decays at long enough times $t>>t_\times$
Indeed, as was shown in the inset of Fig.~\ref{fig:cg-c-d00}, 
such a hump-like weak structure of the spin autocorrelation 
was 
discernible in our present data of $C_s(t)$ 
at short times $t\simeq 10^2$, which, however,
eventually decayed toward zero at longer times. 

Berthier and Young also observed  in their recent
{\it off-equilibrium\/}
simulation of the 3D Heisenberg SG a weak
hump-like structure in the spin autocorrelation in the time range 
$t\alt 10^4$,
which corresponded to the quasi-equilibrium regime \cite{BY}. 
These authors interpreted the observed hump
as an evidence of a nonzero SG long-range order at that 
temperature. 
As noted above, however, such a hump is also consistent with the
the spin-chirality coupling/decoupling picture as long as the hump 
is observed only at shorter times $t<t_\times$.

\subsection{Nonequilibrium dynamics}

Nakamura and Endoh applied a non-equilibrium method to study the
SG and the CG orderings of the 3D $\pm J$ Heisenberg SG.\cite{Nakamura} 
Analyzing the time dependence of the initial growth of the SG and the 
CG susceptibilities with use of a 
dynamical scaling,  these authors concluded that the spin
and the chirality ordered  simultaneously at a finite temperature
$T/J=0.21\sim 0.22$. While
the lattice size studied $L\leq 59$ was rather large,
the crucial question to be addressed is whether the long-time limit 
$t\rightarrow \infty$  was safely  taken justifying 
the use of the dynamical scaling. 
In other words,  although the nominal
lattice size studied was large, 
the equilibrated length scale actually probed in these off-equilibrium
simulations might
be rather short. In fact, their non-equilibrium method is uncontrolled
with respect to the time scale toward equilibrium. Since 
the equilibration time could easily become a huge number in SG,
care
has to be taken 
as regards the equilibrated length scale actually probed by
this type of non-equilibrium simulation. 
As one judges from the maximum values
of the SG and the CG susceptibilities reached by 
the off-equilibrium 
simulation of Ref.~\onlinecite{Nakamura}, the ``dynamical correlation
length'' still remained rather short: Namely,
even around the transition 
temperature $T_{{\rm CG}}/J\simeq 0.2$, 
it reached around 10 lattice spacings for the spin,
and only 
1 or 2 lattice spacings for the chirality. The dynamical
chiral correlation length stayed particularly short. 
This is consistent with 
a recent off-equilibrium simulation by Berthier and Young in which 
the dynamical chiral correlation length stayed much shorter than the
dynamical spin correlation length in the investigated time range \cite{BY}. 
Here note that, irrespective of the question of whether the 
CG transition accompanies the simultaneous SG
transition or not, the chiral correlation length in equilibrium should diverge 
at and below the CG transition temperature 
$T_{{\rm CG}}/J\simeq 0.2$. 
Hence, the observation above simply tells that,
even at the maximum simulation time of Refs.\onlinecite{Nakamura} and
\onlinecite{BY},   
the system still stayed in an extreme initial time regime. In order to deduce
the equilibrium ordering properties from these off-equilibrium data, 
one is forced to extrapolate the behavior around 
$\xi_{{\rm CG}} \sim 1$ to $\xi_{{\rm CG}} =\infty$, 
which could be dangerous in  
the present model since the model might possess 
the characteristic crossover length
scale at around 20.

One may feel that the dynamical spin correlation length 
reached in the off-equilibrium simulation of Ref.~\onlinecite{Nakamura}, 
$\xi \approx 10$, 
might be reasonably large for deducing the ordering properties of the spin. 
However, we feel it is not enough. 
This length scale of 10 is still not
large enough compared with the crossover length scale estimated in the
present work
$L_\times \approx 20$. Remember that the spin-chirality decoupling, if any,
is a long scale phenomena observable at length scale longer than $L_\times$. 
Second, in the off-equilibrium simulations of
Refs.~\onlinecite{Nakamura} and \onlinecite{BY}, even  when
the dynamical
SG correlation length grows around 10 lattice spacings,
the $Z_2$ chiral degree of freedom was not equilibrated at all
at this length scale,
in sharp contrast to the fully equilibrated simulation as was done in the
present paper.
In other words, at the length scale of 10 lattice spacings,
the chiralities are little thermalized and are virtually frozen in the
non-equilibrium pattern, while only the SG correlation  
grows modestly in such a 
nonequilibrium chiral environment. After all, however, we have to understand
the spin dynamics at long enough length scales
at which the $Z_2$ chiral degree of freedom is also fully 
thermalized. Thus, 
the spin dynamics as 
observed in the off-equilibrium situations of Refs.~\onlinecite{Nakamura}
and \onlinecite{BY} may not faithfully represent the close-to-equilibrium critical 
dynamics of the original model. 

A similar dynamical simulation on the 3D $\pm J$ 
Heisenberg SG 
was performed by Matsumoto, Hukushima and Takayama.\cite{Matsumoto}  
They also made a dynamical 
scaling analysis taking the effect of global spin rotations into account. 
In this study, 
the time scale toward equilibrium was controlled via the analysis of 
the waiting-time dependence of the results.
In contrast to Refs.~\onlinecite{Nakamura} and \onlinecite{BY},
Matsumoto {\it et al\/} 
suggested that their
data were consistent with the 
separate spin and chiral transition, {\it i.e.\/},
$T_{{\rm SG}}<T_{{\rm CG}}$.

Berthier and Young also argued that their observation that the 
dynamical CG correlation length stayed shorter than the dynamical SG
correlation length presented 
an evidence that the spin, rather than the chirality,
was the order parameter of the transition.\cite{BY} 
Some caution is required in drawing the final conclusion from this
observation, though.
Both the spin and the chirality length scales probed by 
the off-equilibrium simulations of Refs.\onlinecite{Nakamura} and \onlinecite{BY}
are still shorter than the
crossover length estimated in the present paper $L_\times 
\simeq 20$. Hence,
within the spin-chirality coupling/decoupling scenario,
there still exists
a good possibility that the dynamical CG correlation length 
eventually outgrows 
the dynamical SG correlation length at longer times.

In Ref.~\onlinecite{BY}, the aging phenomena were persistently 
observed at lower temperatures, 
not only for the chirality but also for the spin, 
which was interpreted
as an evidence of the occurrence of simultaneous spin and chiral
transition.\cite{BY}  
Again, this cannot be taken as an unambiguous indicator
of a finite-temperature SG transition, since 
the aging phenomena could arise simply when the
time scale of measurements becomes comparable to
the longest relaxation time in the system 
which could be extremely long in
SG even in the paramagnetic phase. For example, in the 2D Ising
SG which is known to exhibit no finite-temperature SG transition,  clear aging
phenomena have been observed 
both in numerical simulations \cite{Svedlindh,Rieger} and 
in experiments \cite{Hammann,Nordblad}.

\subsection{Correlation length}

In Ref.~\onlinecite{LeeYoung}, 
Lee and Young calculated by means of equilibrium simulations 
both the SG and the CG correlation lengths
of the 3D Heisenberg SG with the Gaussian coupling in the range of
sizes $4\leq L \leq 12$. Lee and Young observed a crossing of the 
dimensionless correlation lengths $\xi/L$ for
different $L$ for both cases of the spin and the chirality, and concluded
that the spin and the chirality ordered simultaneously
at a finite temperature $T/J\simeq 0.15$. 
The behavior of $\xi/L$ observed in Ref.~\onlinecite{LeeYoung} 
turned out to be quite different from
that of some other dimensionless quantities, 
{\it e.g.\/}, the Binder ratio, whereas Lee and
Young argued that the correlation length might be the most trustable quantity
to look at. Generally speaking, however,
$\xi/L$ is also subject to significant finite-size effects, sometimes
no better than other quantities.\cite{KC} 

We note that 
the numerical data of Ref.~\onlinecite{LeeYoung} are basically consistent with
our present data for smaller sizes $L\leq 12$
: See Fig.~\ref{fig:corr-len}(a).
As emphasized in Sec.~\ref{subsec:clength} of our present paper, however,
the crossing
behavior of the dimensionless SG
correlation length $\xi_{{\rm SG}}/L$ tends to change 
for larger lattices $L>12$: 
The crossing becomes weaker and weaker,
and $\xi_{{\rm SG}}/L$
of $L=16$ and that of $L=20$ do not quite cross with a finite crossing angle as
occurs for smaller lattices $L\leq 12$, but instead, merge almost
tangentially and stay on top of each other at lower temperatures: 
See Fig.~\ref{fig:corr-len}(b). In contrast,
the dimensionless CG correlation length
$\xi_{{\rm CG}}/L$ of $L=16$ and that of $L=20$ persistently exhibits 
a clear crossing. If such a tendency continues for  larger
lattices, 
the crossing of $\xi_{{\rm SG}}/L$ might no longer
occur for large enough lattices, at least
at the crossing temperature of $\xi_{{\rm CG}}/L$.
We thus suspect that the crossing behavior of $\xi_{{\rm SG}}/L$ as reported
in Ref.~\onlinecite{LeeYoung} might be a transient behavior due to the small sizes, 
which reflected  the
``coupling'' behavior expected at $L<L_\times \approx 20$.
Namely, within the  spin-chirality coupling/decoupling scheme, 
the SG
correlation exhibits the ordering behavior similar to the CG correlation
at shorter length scales at which the spin is coupled to the chirality, 
while at longer length scales at which 
the spin is decoupled from the chirality,
the SG correlation eventually exhibits the non-ordering behavior 
different from the CG correlation.
Unfortunately,
the largest lattice size accessible by the
present computational capability $L\simeq 20$,
being only comparable to the
crossover length for the spin-chirality coupling/decoupling to occur, 
is still not large enough to
clear see this behavior. 
We do expect, however, 
that the correlation lengths 
for larger lattices $L>20$ would eventually
exhibit a clear spin-chirality decoupling behavior.

We note that such a coupling/decoupling behavior of the
SG correlation length in smaller/larger lattices
was indeed observed recently in the 2D Heisenberg SG.\cite{KawaYone}
For the 2D Heisenberg SG with the Gaussian coupling, Kawamura and Yonehara
calculated the dimensionless SG correlation length
$\xi_{{\rm SG}}/L$ up to the size $L$=40, and found that 
$\xi_{{\rm SG}}/L$ for the smaller sizes $L=10,16,20$
crossed almost at a common temperature $T/J\simeq 0.022$,
disguising the occurrence of a finite-temperature 
SG transition: See the inset of
Fig.6(a) of Ref.~\onlinecite{KawaYone}, while the data for the larger sizes
$L$=30 and 40 data eventually came down, no longer making a crossing
at $T/J\simeq 0.022$. The asymptotic non-ordering
behavior observed for $L>20$ is
consistent with a zero-temperature SG transition, which has been  well
established in 2D.\cite{KawaYone,Schwartz}  
Meanwhile, the CG correlation length exceeds the
SG correlation length at around $T/J\simeq 0.022$,
which might naturally explain the reason 
why $\xi_{{\rm SG}}/L$  
for smaller sizes $L\alt 20$ exhibited a crossing behavior.
Anyway, this observation in 2D
gives us a warning that one should be careful in interpreting
the crossing-like  behavior of $\xi/L$ observed for smaller sizes 
as an unambiguous evidence of a true SG phase transition.

\subsection{Finite-size scaling of the order parameter}

Matsubara, Shirakura, Endoh and Takahashi 
made a finite-size scaling analysis of 
the SG order parameter $q^{(2)}_{{\rm SG}}$ for the
3D $\pm J$ Heisenberg SG,  and claimed that the
quality of the scaling was much better when ones assumed a nonzero SG
transition temperature $T_{{\rm SG}}/J=0.18$ than
a zero SG transition temperature $T_{{\rm SG}}/J=0$.\cite{Matsubara3} 
Their conclusion is in apparent contrast to that of our present work 
based on a similar scaling analysis in Sec.~\ref{subsec:fss-q2}.   
We note that 
the quality of the finite-size scaling is sometimes sensitive to the range
of lattice sizes and the range of temperatures used in the fit.

As already noticed in Sec.~\ref{subsec:fss-q2}, 
this point could be particularly serious in the present model.
In the spin-chirality coupling/decoupling scheme,  the SG correlation 
and the CG correlation are trivially coupled at shorter length scale
$L\alt L_\times \approx 20$, 
so that even the SG order parameter $q^{(2)}_{{\rm SG}}$
would be scaled for smaller sizes
with assuming a simultaneous SG and CG transition, with  
apparent (not true) SG pseudo-exponents 
$\nu^{\rm eff}_{{\rm SG}}\simeq \nu_{{\rm CG}}$ and 
$1+\eta^{\rm eff}_{{\rm SG}} \simeq (1+\eta_{{\rm CG}})/3$. 
Indeed, as was shown 
in Fig.~\ref{fig:fss-q-d00-2},
our present data,  particularly those of $L\alt 16$, turned out to
be scaled reasonably well by assuming a simultaneous SG and CG
transition at $T/J=0.19$, which we interpreted as a pre-asymptotic 
pseudo-critical behavior realized in the short-scale coupling regime.
Furthermore, the relation between $\eta_{\rm CG}$ and
$\eta^{\rm eff}_{\rm SG}$  mentioned above roughly holds at short length scale;
$(1+\eta_{\rm CG})/3\sim 0.60$ versus $1+\eta^{\rm eff}_{\rm SG}\sim
0.88$ at $T/J=0.19$.
At longer length scales, however, 
the spin is eventually decoupled from the chirality. Then, 
if ones continues to put
$T_{{\rm SG}}=T_{{\rm CG}}$ in the fit of $q^{(2)}_{{\rm SG}}$, 
the good data collapse obtained for smaller sizes 
would eventually deteriorate
for larger sizes. Indeed,
as shown in Fig.~\ref{fig:fss-q-d00-2}, 
our $L=20$ data of $q^{(2)}_{{\rm SG}}$ 
showed such a deviation expected for larger sizes.

Within the spin-chirality coupling/decoupling scheme, 
in order to see the true asymptotic 
critical behavior of the SG transition occurring at
$T_{{\rm SG}} (<T_{{\rm CG}}$), 
one has to enter into the long-scale decoupling regime and 
well below the CG
transition temperature, {\it i.e.\/},
$L\geq L_\times \approx 20$ and $T<T_{{\rm CG}}\approx 0.2J$. 
It should be noticed that,
in their scaling fit of $q^{(2)}_{{\rm SG}}$,
Matsubara {\it et al\/} included the data
points
for smaller sizes $L=5,7,9$ {\it etc\/}, 
which are expected to lie in the short-scale
coupling regime, 
as well as the data points at temperatures above $T_{{\rm CG}}$ 
which might lie outside the asymptotic critical regime of the
SG transition. 
Hence, the poor scaling reported by Matsubara {\it et al\/} with
assuming $T_{{\rm SG}}=0$ might simply be due to
the fact that the data points used in the fit 
are not in the correct asymptotic regime.

By contrast, we have observed that, 
if we use the data points only of larger lattices $L\agt 16$
and only at temperatures below $T_{{\rm CG}}$, 
the data were scaled reasonably well
even with assuming $T_{{\rm SG}}=0$: See Fig.~\ref{fig:fss-q-d00}. 
Therefore, we believe that there still
exists a good possibility that the SG order occurs only at $T=0$
as has  widely been believed in the community,
although it is also quite possible that it
occurs at a low but nonzero temperature, $0<T_{{\rm SG}}<T_{{\rm CG}}$.

As discussed in some detail above, any of the recent works claiming the 
simultaneous spin and chiral transition in the 3D Heisenberg SG
appears not conclusive. As far as the authors are aware, 
all of these observations
are consistent with the spin-chirality coupling/decoupling scheme with
the crossover length scale of 20 lattice spacings and the crossover
time scale of $10^5-10^6$ MCS. Rather, we believe that
some of the observations reported in the present paper give a
strong numerical support that  the SG
transition indeed 
occurs at a temperature {\it below\/} the CG transition 
temperature, {\it i.e.\/}, $T_{{\rm SG}}<T_{{\rm CG}}$.

\section{Summary}
\label{sec:summary}

In summary, we studied the equilibrium properties of the three-dimensional 
isotropic Heisenberg spin glass by means of extensive MC
simulations.
We  presented evidence of a finite-temperature CG
transition without accompanying
the conventional SG order through the observation
of various physical quantities
including the order parameters,  equilibrium static and dynamic correlation
functions, Binder parameters and overlap-distribution functions, {\it etc\/}. 
Our conclusion 
is in contrast to some of the recent numerical studies on the same model,
which claimed
the simultaneous SG and CG transition. 
We have pointed out that the crossover length scale
and the crossover time scale associated with the spin-chirality
coupling/decoupling are crucially important in properly interpreting 
the numerical data. 
Around the CG transition temperature, these length and time scales are roughly estimated to be 20 lattice spacings and $10^5\sim
10^6$ MCS, respectively. 
Below these length and time scales, the spin is trivially coupled to the
chirality so that the spin-chirality decoupling, {\it i.e.\/}, the SG 
disorder, is difficult to observe. 
This might give a natural interpretation of the discrepancy between
our present result and the observation of the simultaneous SG 
and CG transition by some other authors. 

Rather, it appears to the authors that our present data for larger $L$
are hard to understand based on the standpoint of the simultaneous
spin and chiral transition. Hence, while simulations on still larger lattices
with $L>20$ are required to settle the issue, our present data
give some support to the spin-chirality decoupling scenario for the
3D isotropic Heisenberg spin glass.

\begin{acknowledgments}   
The authors are thankful to H. Takayama, I. Campbell, H. Katzgraber, M. Picco,
H. Yoshino for valuable discussion and comments.
This work was supported by the Grants-In-Aid (No.~14084204 and No.~16540341) for
 Scientific Research from the Ministry of Education, Culture, Sports,
 Science and Technology of Japan. 
The numerical calculations were mainly performed on 
 the SGI Origin 2800/384 at the Supercomputer Center, Institute for
 Solid State Physics (ISSP), the University at Tokyo.
We would like to thank particularly the ISSP Supercomputer Center for
 granting us exclusive use of the computer system to perform the MC
 simulations presented here. 
\end{acknowledgments}

\end{document}